\documentclass[draftcls,onecolumn,12pt]{IEEEtran}
\pdfoutput=1
\ifCLASSINFOpdf
\else
\fi

\ifCLASSOPTIONcompsoc
\usepackage[caption=false,font=normalsize,labelfont=sf,textfont=sf]{subfig}
\else
\usepackage[caption=false,font=footnotesize]{subfig}
\fi

\usepackage{cite}
\usepackage{bm}
\usepackage{url}
\usepackage[dvipsnames]{xcolor}
\usepackage{amsmath}
\usepackage{mathtools}
\usepackage[]{amssymb}
\usepackage{amsthm}
\usepackage{float}
\usepackage{dsfont}
\usepackage{array}
\usepackage{booktabs}
\usepackage{multirow}
\usepackage{footmisc}

\usepackage{graphicx}
\usepackage[para,flushleft]{threeparttable} 
\usepackage{multirow}	
\usepackage{dcolumn} 

\usepackage[ruled]{algorithm2e}
\usepackage{xifthen} 
\usepackage{enumitem}

\DeclareMathOperator{\diag}{diag}

\def \ba {\begin{array}}
\def \ea {\end{array}}
\def \benu {\begin{enumerate}}
\def \eenu {\end{enumerate}}
\def \bdes {\begin{description}}
\def \edes {\end{description}}

\def \bitem {\begin{itemize}}
\def \eitem {\end{itemize}}

\def \bfl {\begin{flushleft}}
\def \efl {\end{flushleft}}
\def \bfr {\begin{flushright}}
\def \efr {\end{flushright}}

\def \beq {\begin{equation}}
\def \eeq {\end{equation}}
\def \bqa {\begin{eqnarray}}
\def \eqa {\end{eqnarray}}
\def \bqa* {\begin{eqnarray*}}
\def \eqa* {\end{eqnarray*}}

\def \bal {\begin{align}}
\def \eal {\end{align}}

\newcounter{mytempeqncnt}

\begin{document}
%
\title{Frequency and Phase Synchronization in Distributed Antenna Arrays Based on Consensus Averaging and Kalman Filtering}

\author{Mohammed Rashid,~\IEEEmembership{Member,~IEEE}, and Jeffrey A. Nanzer,~\IEEEmembership{Senior Member,~IEEE}

\thanks{Manuscript received 2022.}
\thanks{This work was supported in part by the Office of Naval Research under grant number N00014-20-1-2389. Any opinions, findings, and conclusions or recommendations expressed in this material are those of the author(s) and do not necessarily reflect the views of the Office of Naval Research. \textit{(Corresponding author: Jeffrey A. Nanzer.)}}
\thanks{The authors are with the Electrical and Computer Engineering Department, Michigan State University, East Lansing, MI 48824 (e-mail: \mbox{rashidm4@msu.edu,} nanzer@msu.edu).}}


\maketitle
\thispagestyle{empty}
\pagestyle{empty}

\def\bda{\mathbf{a}}
\def\bdd{\mathbf{d}}
\def\bde{\mathbf{e}}
\def\bdf{\mathbf{f}}
\def\bdg{\mathbf{g}} 
\def\bdh{\mathbf{h}}
\def\bdm{\mathbf{m}}
\def\bds{\mathbf{s}} 
\def\bdn{\mathbf{n}}
\def\bdu{\mathbf{u}}
\def\bdv{\mathbf{v}}
\def\bdw{\mathbf{w}} 
\def\bdx{\mathbf{x}} 
\def\bdy{\mathbf{y}} 
\def\bdz{\mathbf{z}}
\def\bdA{\mathbf{A}}
\def\bdC{\mathbf{C}}
\def\bdD{\mathbf{D}} 
\def\bdF{\mathbf{F}}
\def\bdG{\mathbf{G}} 
\def\bdH{\mathbf{H}}
\def\bdI{\mathbf{I}}
\def\bdJ{\mathbf{J}}
\def\bdX{\mathbf{X}}
\def\bdK{\mathbf{K}}
\def\bdQ{\mathbf{Q}}
\def\bdR{\mathbf{R}}
\def\bdS{\mathbf{S}}
\def\bdV{\mathbf{V}}
\def\bdW{\mathbf{W}}
\def\bdGamma{\bm{\Gamma}}
\def\bdgamma{\bm{\gamma}}
\def\bdalpha{\bm{\alpha}}
\def\bdmu{\bm{\mu}}
\def\bdSigma{\bm{\Sigma}}
\def\bdxi{\bm{\xi}}
\def\bdl{\bm{\ell}}
\def\bdLambda{\bm{\Lambda}}
\def\bdeta{\bm{\eta}}
\def\bdPhi{\bm{\Phi}}
\def\bdtheta{\bm{\theta}}
\def\bddelta{\bm{\delta}}

\def\btau{\bm{\tau}}
\def\deg{\circ}

\def\tq{\tilde{q}}
\def\tbdJ{\tilde \bdJ}
\def\l{\ell}
\def\bdzero{\mathbf{0}} 
\def\bdone{\mathbf{1}} 
\def\Exp{\mathbb{E}} 
\def\exp{\text{exp}} 

\def\R{\mathbb{R}} 
\def\C{\mathbb{C}} 
\def\CN{\mathcal{CN}} 
\def\N{\mathcal{N}} 
\begin{abstract}
A decentralized approach for joint frequency and phase synchronization in distributed antenna arrays is presented.
The nodes in the array 
locally broadcast their frequencies and phases to their neighboring nodes, and use consensus averaging to align these 
parameters across the array. 
The architecture is fully distributed, requiring no centralization.
Each node has a local oscillator and
we consider a signal model where intrinsic frequency and phase errors 
of the local oscillators on each node caused by the frequency drift and phase jitter as well as the 
frequency and phase estimation errors at the nodes are included and modeled 
using practical statistics. 
A decentralized frequency and phase consensus (DFPC) algorithm is proposed 
which uses an average consensus method in which each node in the array iteratively 
updates its frequency and phase by computing an average of the frequencies and phases of 
their neighboring nodes. Simulation results show that upon convergence the DFPC algorithm can align the 
frequencies and phases of all the nodes up to a residual phase error that is governed by 
the oscillators and the estimation errors. To reduce this residual 
phase error and thus improve the synchronization between the nodes, a Kalman filter based 
decentralized frequency and phase consensus (KF-DFPC) algorithm is presented. 
The total residual phase error at the convergence of the KF-DFPC and DFPC algorithms 
is derived theoretically. 
The synchronization performances of these algorithms 
are compared to each other {and to the diffusion least mean square (DLMS), the 
diffusion KF (DKF) algorithm, and the Kalman consensus information filter (KCIF) algorithms,} 
in light of this theoretical residual phase error by varying the duration of the 
signals, connectivity of the nodes, the number of nodes in the array, and signal to noise 
ratio of the received signals. {Simulation results demonstrate that under certain conditions 
the proposed KF-DFPC algorithm outperforms and 
converges in fewer iterations than all the algorithms}. 
Furthermore, for shorter intervals between local information broadcasts,
the KF-DFPC algorithm significantly outperforms the DFPC algorithm 
in reducing the residual total phase error,
irrespective of the signal to noise ratio of the received signals.  
\end{abstract}


\section{Introduction}

Wireless systems evolution from large single-platform 
architectures to a composite of multiple small spatially distributed systems 
that make local decisions to achieve a shared global objective 
has resulted in significant applications in distributed coordinated 
beamforming \cite{DCB_2013, NetBeam_2019}, network densification in 
5G \cite{Densification_5G}, and coherent distributed array (CDA) 
\cite{OCDA_2017, Sean_2020, Sean_TMTT_2019}. 
In particular, CDA operates as a network of small, low-cost, and spatially 
distributed nodes where each node has its own separate transceiver, and the nodes 
align their transceivers within a fraction of the wavelength of the wireless operation 
in such a way that their signals add up coherently at the destination. 
This coordination of the nodes at the level of the wavelength imitates a 
distributed phased array and results in 
several advantages at the overall system level that include 
improved signal-to-noise (SNR) ratio at the destination, enhanced resistance to the system 
failure, greater spatial diversity, and considerable ease in the scalability of the 
system \cite{Nanzer_Survey_2021}.  

In coherent distributed arrays, each node has its own local oscillator in its 
transceiver chain. {In an unlocked state, these oscillators undergo random frequency and phase drifts, and phase 
jitter over time due to the factors such as the temperature variations, noise from the supply voltage, and 
the manufacturing tolerances of its components, etc. \cite{Serge_Access_2021}. 
If the nodes generate their carrier signals by using reference signals from their own local oscillators then 
due to these frequency and phase offsets, there exist a decoherence between their generated signals which 
distorts the coherent gain at the destination. Naively, we can connect these nodes via wires 
to a common reference signal \cite{Wired_sync}, 
but that severally limits the mobility of the nodes as well as the application domain of such arrays. 
Another way is to connect each node to the global positioning system (GPS) signal and use it to synchronize them across the array \cite{GPS_syn}, but the 
GPS-based systems are expensive, highly power consuming, and do not work in the indoor environments. 
Furthermore, wireless locking of the oscillators to a central oscillator have also been investigated using the 
wireless signals in \cite{Mabel_2017, Mabel_2019} and using 
the optical signals in \cite{Yang_2014}. However, these methods are incapable of synchronizing the nodes at 
larger distances and are also not scalable.} 

%
%

To date, several methods have also been proposed for the wirelessly synchronizing of the nodes in a coherent 
distributed array that can be broadly classified as either closed-loop or open-loop 
methods. In the closed-loop methods, such as the 1-bit and 3-bit feedback methods in 
\cite{1_bit_feedback, 3_bit_feedback, Seo_2008, Quitin_2013} and the retrodirective method in \cite{retrodirective_2016}, 
the nodes adjust their electrical states based on the feedback from the destination. This feedback 
is in terms of some useful information, e.g., 
received signal strengths, preambles, or data throughput, etc., and the nodes use 
the feedback to adjust their states with the intent to reach a desirable coherent gain 
at the destination. 
However, closed-loop methods are only suitable for scenarios where such meaningful 
feedback is available from the destination, for instance, in communication applications, 
thus limiting its ability to arbitrarily steer the beam to any destination. 
On the other hand, in the open-loop methods, such as the synchronization 
methods in \cite{Hassna_TWC, Hassna_TAP, Schmid_2017, Serge_2021}, the nodes 
coordinate with each other to synchronize their transceivers without using any 
feedback from the destination. As such, open-loop methods are also suitable 
for remote sensing \cite{SAR_tutorial, distd_SAR} and radar applications 
\cite{Schmid_radar}.

For open-loop CDAs, a centralized topology based transceiver synchronization 
method is proposed in \cite{Serge_2021, Serge_2019, AirShare_2015} where a primary node transmits 
a reference signal to one or more secondary nodes, and the secondary nodes 
use a frequency locking circuit with phase-locked loops to 
synchronize their frequencies with the frequency of the primary node. 
A drawback of this primary-secondary architecture is that it fails whenever the primary node fails. 
Thus the authors in \cite{Hassna_TWC, Hassna_TAP} proposed a decentralized 
algorithm for nodes synchronization in open-loop CDAs. The decentralized topology 
overcomes the shortcomings of the centralized one, and makes the system easily scalable.

{One limitation of \cite{Hassna_TWC, Hassna_TAP} is that it only considered frequency synchronization of the 
oscillators in distributed arrays, however, in practice, the phases of the oscillators also undergo 
random drifts and jitters over time and thus need to be aligned as well 
for achieving high coherent gain at the destination \cite{Serge_MWCL_2022, Serge_Access_2021}. 
Consequently, we consider the joint frequency 
and phase synchronization of the nodes in a distributed antenna array, and to this end,  
the contributions made in this paper are summarized as follows. 
\begin{itemize}
\item We consider a signal model for the nodes in which the oscillator frequency and phase drifts 
as well as the phase jitters are included to demonstrate a more practical scenario. 
Furthermore, due to the addition of these random offsets, we assume that the nodes iteratively estimate their frequencies and phases before updating them, 
and thus the estimation errors are also included in the signal model. An iterative decentralized 
frequency and phase consensus (DFPC) algorithm is proposed which synchronizes the nodes across the array 
in a few iterations. 
This joint frequency and phase synchronization was also studied in 
\cite{Mallada_2011, Mallada_2016}, but their proposed algorithm synchronizes the oscillators 
to the harmonic means of their initial frequencies \cite{Mallada_2011}. On the contrary, 
to average out the offset errors and thereby reduce the residual phase errors, 
it is shown in Section \ref{steady_state_section} in this paper that 
the convergence to the arithmetic average value is essential in the presence of the 
frequency drifts and phase jitters of the oscillators, which is ensured by DFPC. 
\item Steady-state residual total phase error of the DFPC algorithm 
is also theoretically derived herein by taking into account the frequency and phase 
drifts, phase jitters, and frequency and phase estimation errors at the nodes. 
It is observed that the residual phase error of DFPC decreases with the increase 
in either the number of nodes in 
the array or the connectivity between the nodes. This is also illustrated through simulation in Section 
\ref{residual_phase_section} where the phase errors are analyzed for different arrays 
by varying the update intervals and the signal-to-noise ratio (SNR) of the signals.
\item In practice, usually the frequencies and phases of the nodes have to be updated periodically with 
smaller update intervals to avoid the decoherence between the nodes caused by 
the large oscillators drift in longer time intervals, and to minimize the system impacts 
due to the platform vibrations \cite{Pratik_2017}. However, 
Section \ref{residual_phase_section} shows that the DFPC algorithm 
results in larger residual phase errors at the smaller update intervals. Similar behavior was 
also observed for the frequency synchronization algorithms proposed in \cite{Hassna_TWC, Hassna_TAP} which 
is caused by the increase in the estimation errors. 
Kalman filtering (KF) is a popular method that has been used in the past in a variety of problems 
ranging from target tracking to the industrial control \cite{Urrea_2021}. 
Essentially, it computes the optimal minimum 
mean squared error estimate of the unknown quantities when the 
process noise in the state transitioning model and the measurement noise of the observations 
are normally distributed, and the observations 
are a linear function of the unknown states \cite{Anderson1979}. To take advantage of this KF property, 
we integrate KF with DFPC to reduce the residual phase errors at the shorter update intervals where 
the measurement errors are dominant and the states are slowly time-varying. 
The resulting algorithm is referred to as the KF-DFPC algorithm. Simulation results 
are included where the improvement due to KF at the smaller update intervals is illustrated and compared to DFPC. 
Furthermore, the synchronization performance and the computational complexities of KF-DFPC and DFPC are compared to 
the earlier proposed diffusion least mean square (LMS) algorithm \cite{LMS_2010}, the Kalman consensus information 
filter (KCIF) algorithm \cite{Saber_2009}, and the diffusion KF algorithm \cite{Sayed_2010, Xin_2022}. The results show that 
our proposed KF-DFPC converges faster at lower SNRs and when sparsely connected arrays are used, whereas for the 
large densely connected arrays, the improvement with using KF 
comes with no additional increase in the computational complexity of the DFPC algorithm.
\end{itemize}
}

The upcoming sections are outlined as follows. Section \ref{Freq_Phase_OCDA} introduces the 
frequency and phase synchronization problem in a distributed antenna array, and describes the modeling of the 
frequency and phase errors. Section \ref{DFPC_section} proposes the DFPC algorithm, 
derives the steady-state total phase error, and analyzes DFPC performance through simulations. 
To reduce the residual total phase error and improve the synchronization between nodes, 
Kalman filtering based KF-DFPC algorithm 
is proposed in Section \ref{KF_freq_phase_sync} wherein it is also studied through simulations. 
Section \ref{residual_phase_section} analyzes the residual phase errors of DFPC and KF-DFPC for different update intervals 
and connectivity values. 
Finally, Section \ref{conclusion_section} concludes this work.

\textit{\textbf{Notations:}} Herein, small letters $(x)$ are used to 
represent scalars or signals, bold small letters $(\bdx)$ are used for vectors, 
and bold capital letters $(\bdX)$ are used for matrices. 
The superscripts $(.)^T$ and $(.)^{-1}$ represent the matrix transpose and inverse 
operations, respectively. $\N(\bdmu,\bdSigma)$ denotes a normal distribution with 
mean $\bdmu$ and covariance matrix $\bdSigma$. $\bdI_N$ denotes the $N \times N$ identity 
matrix. In $\Exp[\bdx]$, $\Exp[.]$ denotes the expectation operation with respect to the 
probability distribution on $\bdx$. $\bdx^{1:k}_n$ is a shorthand form for the set 
$\left\{\bdx_n(1),\bdx_n(2),\ldots,\bdx_n(k)\right\}$. Finally, $\bdX=\diag\{\bdx\}$ 
is a diagonal matrix formed by the elements of vector $\bdx$ on its main diagonal.  

\section{Frequency and Phase Synchronization in Open-Loop Coherent Distributed Antenna Arrays}\label{Freq_Phase_OCDA}

In open-loop coherent distributed arrays, the nodes synchronize their transceivers in frequency, 
phase, and time by coordinating with each other (without any feedback from the destination) to 
perform a shared coherent operation. Each node derives its carrier frequency from its own local 
oscillator in the transceiver chain. An oscillator undergoes a random frequency drift over time 
that offsets its carrier frequency, resulting in decoherence of the signals emitted by the nodes.
Thus frequency synchronization of the nodes is crucial to avoid the incoherence at the destination 
due to the frequency offset. 
Furthermore, the phase of an oscillator also undergoes random drift and jitter over time 
that offsets the carrier phase. 
If the signals from the nodes are not synchronized in phase, they may add destructively 
at the destination resulting in a degradation of the coherent gain. Time synchronization is also needed so that 
all the nodes can perform the coherent operations within the same coordinated interval of the clock. 
Note that the errors due to loss in frequency, phase, and time synchronization are independent 
and can be corrected independently and analyzed in the total error budget \cite{OCDA_2017}. 
For open-loop CDAs, time synchronization is studied in \cite{Nanzer_Survey_2021, Pratik_2018} and 
thus we assume herein that the nodes have been synchronized 
in time to the desired level. In this 
work, we consider the problem of frequency and phase synchronization of the nodes in the system 
and thus present a decentralized approach for their joint synchronization. 

\subsection{Signal Model with Frequency and Phase Errors}\label{modeling_section}
 
Consider a network of $N$ nodes spatially separated by multiples of the wavelength of 
coherent operation and coordinating with each other to form a coherent beam in a 
given direction. The signal received in their mutual far-field 
can be given by
\begin{align}\label{Sig_model}
s(t)&=C \sum^N_{n=1} e^{j\left(2\pi f_c t + \frac{2\pi}{\lambda_c}d_n 
\text{cos}(\theta_n)+\delta \phi_n \right)}, 
\end{align}
where $C$ is a constant amplitude term, $f_c$ is the carrier frequency of the coherent operation 
and $\lambda_c$ is its wavelength, 
$d_n$ and $\theta_n$ are the distance and orientation angle of 
the $n$-th node relative to a reference point or node, and $\delta \phi_n$ represents the total 
phase error at the $n$-th node. The phase error $\delta \phi_n$ is defined as 
$\delta \phi_n= 2 \pi \delta f_n T+2 \pi \varepsilon_f T+\theta^e_n +\delta \theta^f_n+\delta \theta_n+\varepsilon_\theta$ 
in which $2\pi \delta f_n T$ is the phase error due to the frequency offset $\delta f_n$ 
of the oscillator in the $n$-th node at the update time $T$; 
the term $2 \pi \varepsilon_f T$ 
represents the phase due to frequency estimation error $\varepsilon_f$ at the $n$-th node assuming that 
the nodes need to estimate their oscillator frequency in the update interval to synchronizing it 
with the other nodes; $\theta^e_n$ is a constant phase term assumed herein to be uniformly distributed between 
$0$ to $2\pi$ and accounting for the estimation error in the distance 
and orientation angle of the $n$-th node, any residual phase error due to the timing misalignment, 
any hardware or channel induced phase delay, and any phase error due to the mismatch of antennas radiation 
pattern, etc.; $\delta \theta^f_n$ represents the phase error 
due to the time variation of the frequency of the oscillator within the update interval; 
$\delta \theta_n$ is the error due to the random phase jitter of the oscillator 
in the $n$-th node; and $\varepsilon_\theta$ is the phase estimation error at the $n$-th node assuming 
that the nodes must estimate their oscillator phase in the update interval to synchronize it with the other nodes 
\cite{OCDA_2017, Hassna_TWC, Serge_Access_2021}. Note that for an ideal coherent operation, 
the total phase error $\delta \phi_n$ must be zero, however, it has been shown in \cite{OCDA_2017} that 
at least $90 \%$ of the ideal coherent gain can be achieved at the destination if this total phase 
error is below the $18^\deg$ threshold (see Fig. 4 in \cite{OCDA_2017}).

\subsection{Modeling Frequency and Phase Errors}\label{modeling_ftheta}

Herein we discuss the modeling of the errors 
$\{\delta f_n, \delta \theta^f_n, \delta \theta_n, \varepsilon_f, \varepsilon_\theta\}$ 
for all the nodes, i.e., $n=1,2,\ldots, N$ as follow. 

The frequency stability of an oscillator is characterized mainly by its design metrics which includes 
the manufacturing tolerances of its components and their susceptibility to temperature variations. 
Allan deviation (ADEV) is a popular metric which is used to quantify the frequency drift $\delta f_n$ 
of an oscillator in the $n$-th node. 
It is computed by averaging the fractional frequency error over multiple shifted time 
intervals and then computing the standard deviation of these measurements. 
The ADEV value is usually time varying such that in shorter time 
intervals its value is governed 
mainly by the noise fluctuations, but as the time passes, these fluctuations equalize resulting in a 
smaller ADEV and a stable oscillator frequency \cite{Serge_Access_2021}. 
For a temperature compensated crystal oscillator, the ADEV value usually ranges from $10^{-9}$ to $10^{-10}$ 
at $T=1$ sec interval. 
Following \cite{Hassna_TWC}, the ADEV can be modeled as 
\begin{equation}\label{ADEV_Eqn}
 \sigma_f=f_c \sqrt{\frac{\beta_1}{T}+\beta_2 T},
\end{equation}
where $\beta_1$ and $\beta_2$ depend on the design of an oscillator and $T$ is the frequency update interval. 
In this work we set $\beta_1=\beta_2=5 \times 10^{-19}$ {to model a quartz crystal oscillator~\cite{Hassna_TWC}} 
and model the frequency drift $\delta f_n$ of an oscillator 
in the $n$-th node as normally distributed with zero mean and standard deviation given 
by \eqref{ADEV_Eqn} \cite{Hassna_TWC}. 

The time variation of ADEV metric as discussed above implies that the frequency offset is also 
time varying. Thus the phase term $\delta \theta^f_n$ in $\delta \phi_n$ in \eqref{Sig_model} 
denotes the phase adjustment needed at the update time $T$ by taking into account 
the variation of the frequency offset over time. 
As discussed in \cite{Serge_Access_2021}, the instantaneous frequency offset can be modeled on 
average as varying linearly over the update interval $T$ as a function $\frac{\delta f_n}{T} t$. 
Hence, the actual phase at time $T$ is calculated as 
\begin{align}
\delta \theta_{\text{actual}}=2\pi \int^T_0 \left(f_c+\frac{\delta f_n}{T}t\right) dt,
\end{align}
and the phase adjustment $\delta \theta^f_n$ at time $T$ is then given by
\begin{align}\label{theta_f_n}
\delta \theta^f_n=\delta \theta_{\text{actual}}-2\pi \int^T_0 \left(f_c+\delta f_n\right) dt,
\end{align}

Now the signal generated by an oscillator in practice is also corrupted by its phase noise which is a 
combination of interference from the external noise sources that includes noise from the power supply and bias 
currents in MOSFETs and BJTs in the circuit, and random fluctuation of the phase 
generated internally in the oscillator. 
The phase noise varies randomly in the oscillator and causes a phase jitter in its generated signal. 
This phase jitter can be measured from the phase noise profile of an oscillator which can be created 
by dividing the possible range of offset frequency into different operating regions and measuring the 
noise contribution due to frequency random walk, frequency/phase flicker, and white frequency/phase 
in each region (see Fig. 3 in \cite{Serge_Access_2021}). The phase jitter is a function of the 
integrated phase noise power $A$ which is computed from the log of the sum of the 
areas of all the regions in the phase noise profile of an oscillator. The phase jitter is defined as 
\begin{equation}\label{Jitter_Eqn}
 \sigma_\theta=\sqrt{2\times 10^{A/10}}
\end{equation} 
Several mathematical models exist for modeling the phase noise profile of oscillators 
\cite{Leeson_1966, Phase_noise_2000}, and typically the value of $A$ ranges from $-103.05$ dB to 
$-53.46$ dB for low to high phase noise oscillator, respectively. 
In this work, we choose $A=-53.46$ dB $\left(\text{i.e., } \sigma_\theta=2.7\times 10^{-3}\right)$ 
that defines the phase noise of a typical voltage controlled 
oscillator. Thus the phase term $\delta \theta_n$ modeling the phase jitter in 
$\delta \phi_n$ from \eqref{Sig_model} is assumed as normally distributed with zero mean and 
standard deviation given by \eqref{Jitter_Eqn} \cite{Serge_Access_2021}. 

Finally, we model the frequency and phase estimation errors, i.e., $\varepsilon_f$ and $\varepsilon_\theta$, 
for all the nodes in a distributed antenna array as follows. 
Since both frequency and phase of an oscillator in the 
node's transceiver are influenced by the 
random offsets, their exact values are unknown to the nodes and must be estimated. Furthermore, 
the nodes may enable a local broadcast to share their frequencies and phases with 
their neighboring nodes. 
To estimate these parameters, we assume that each node collects $L=Tf_s$ samples of 
its signals over an an observation window of length $T$ 
with sampling frequency $f_s$, and uses, for instance, fast Fourier transform based maximum likelihood 
estimators as discussed in \cite{Liao2011,Zhou_2019}. 
The standard deviation of the frequency and phase estimation errors is 
lower bounded by the Cramer-Rao lower bound (CRLB) equations 
which are derived in \cite{Richards_radar}. The CRLB for the frequency estimation error is given by
\begin{equation}\label{f_est_eqn}
\sigma^m_f\geq\sqrt{\frac{6}{(2\pi)^2 L^3 \text{SNR}}}
\end{equation}
and the CRLB for phase estimation error is 
\begin{equation}\label{theta_est_eqn}
\sigma^m_\theta\geq \frac{2L^{-1}}{ \text{SNR}}
\end{equation}

As the focus of this work is on the frequency and phase synchronization rather than designing the estimator, 
we model $\varepsilon_f$ and $\varepsilon_\theta$ as normally distributed with zero mean 
and standard deviation given by the CRLBs in \eqref{f_est_eqn} and \eqref{theta_est_eqn}, respectively. 
In practice, these standard deviations can be replaced by the standard deviation of the estimator used 
for frequency and phase estimation.



\section{Decentralized Frequency and Phase Synchronization}\label{DFPC_section}

We model the network of $N$ nodes in a distributed antenna array by an undirected graph 
$\mathcal{G}=(\mathcal{V},\mathcal{E})$ where 
$\mathcal{V}=\{1,2,\ldots,N\}$ represents the set of vertices (nodes) and 
$\mathcal{E}=\{(i,j)\colon i,j\in \mathcal{V}\}$ is the set of undirected edges (bidirectional 
communication links) between the vertices.  
The signal generated by the $n$-th node in the network is given by 
$s_n(t)=e^{j\left(2\pi f_n t+ \theta_n \right)}$ where $f_n$ and $\theta_n$ represent the 
frequency and phase of its oscillator, respectively.
In the following, we describe a decentralized frequency and phase consensus (DFPC) 
algorithm that iteratively updates the frequency and phase of each node with the aim to synchronize 
these parameters across the array to 
the arithmetic averages of the frequencies and phases of all the nodes in the array. 
To this end, we assume that at iteration $k=0$, each node has an initial estimate of the frequency 
and phase of its signal. 
For the $n$-th node, the initial frequency is selected as 
$f_n(0)=f_c+\mathcal{N}(0,\sigma^2)$ in which $\sigma=10^{-4}f_c$ represents the crystal clock 
accuracy of $100$ parts per million (ppm), and the initial phase is assumed 
to be uniformly distributed between $0$ to $2\pi$, i.e., $\theta_n(0)\sim \mathcal{U}(0,2\pi)$. 
Now let $\bdf(k-1)\triangleq[f_1(k-1),f_2(k-1),\ldots, f_N(k-1)]^T$ and 
$\bdtheta(k-1)\triangleq [\theta_1(k-1),\theta_2(k-1),\ldots,\theta_N(k-1)]^T$ represent the combined 
frequencies and phases of all the nodes in the $(k-1)$-st iteration of the DFPC algorithm, then in 
the $k$-th iteration the algorithm updates the frequencies and phases of all the nodes by  
\begin{align}\label{AC_eqns}
\bdf(k)&=\bdW \bdf(k-1)\nonumber\\
\bdtheta(k)&=\bdW \bdtheta(k-1),
\end{align}
in which $\bdW$ represents the $N\times N$ mixing matrix where its $(i,j)$-th 
element $w_{i,j}$ denotes the weight corresponding to the edge between node $i$ and $j$. 
It is assumed that the matrix $\bdW$ is symmetric and doubly-stochastic (each row and each 
column sums to $1$). Furthermore, its element $w_{i,j}=0$ if nodes $i$ and $j$ are not connected, i.e., 
$(i,j)\notin \mathcal{E}$. This latter property enables decentralized (distributed) averaging of the parameters 
across the network \cite{AC_2004, EXTRA_2015}. 
In this work, we model $\bdW$ as a Metropolis-Hastings matrix \cite{Hassna_TWC, DAC_2007} with $(i,j)$-th element 
defined as 
\begin{equation}\label{mix_W}
w_{i,j}=\left\{
\begin{matrix}
\frac{1}{\text{max} \{n_i,n_j\}+1}, && \text{if } (i,j)\in \mathcal{E}\\
0, && \text{if } (i,j)\notin \mathcal{E} \text{ and } i\neq j\\
1-\sum_{j:j\neq i} w_{i,j}, && \text{if } i= j,
\end{matrix}\right.
\end{equation}
where $n_i$ and $n_j$ represents the number of edges connected to node $i$ and $j$ respectively. 
Note that setting $w_{i,j}=0$ whenever $i\neq j$ in \eqref{mix_W} enables the decentralized property 
across the network where only the local set of weights is needed at each node to compute the average.
It is well known that the modulus of the second largest eigenvalue ($\lambda_2$) of the mixing matrix 
$\bdW$ dictates the 
convergence speed of the consensus algorithm in \eqref{AC_eqns}. An optimization method is proposed 
in \cite{Fast_MC_2004, AC_2004} to design the matrix such that $\lambda_2$ 
can be minimized for faster convergence, however, the method needs the global connectivity information 
of the network which is not feasible in our considered problem.


The nodes in the network are declared to have reached a consensus (synchronization in frequency and phase) 
{when the standard deviation 
of the total phase errors $\delta \phi_n$s' from \eqref{Sig_model} 
is less than or equal to some pre-set threshold $\eta$, i.e.,
\begin{align}
\sigma_\phi&=\sqrt{\frac{1}{N-1}\sum^N_{n=1} \mid \delta \phi_n-\bar{\phi}\mid^2}\leq \eta,
\end{align}
where $\bar{\phi}$ represents the average value of the phase errors.} 

The proposed DFPC algorithm is described in detail in Algorithm \ref{algo_1} 
where the frequency and phase errors discussed earlier in 
Section \ref{modeling_ftheta} are also included in the update process.
\begin{algorithm}\label{algo_1}
  \footnotesize
\DontPrintSemicolon
\SetKwInput{KwPara}{Input}
\KwPara{$k=0$, $\bdW$, $\bdf(0)$, $\bdtheta(0)$.}
\tcc{DFPC run}
\While{convergence criterion is not met} 
{
	$k=k+1$
\begin{enumerate}
\item Define $\bdf(k-1)=\bdf(k-1)+\bddelta \bdf$ where frequency drifts\\ are modeled as 
$\delta \bdf\sim \mathcal{N}\left(\bdzero,\sigma^2_f\bdI_N\right)$.

\item Set $\bdtheta(k-1)=\bdtheta(k-1)+\bddelta \bdtheta^f+\bddelta \bdtheta$ 
where phase errors 
$\bddelta \bdtheta^f\triangleq\left[\delta \theta^f_1,\delta \theta^f_2, \ldots, \delta \theta^f_N\right]^T$ 
are computed from \eqref{theta_f_n}\\ 
and phase jitters are modeled as 
$\bddelta \bdtheta\sim \N \left(\bdzero,\sigma^2_\theta \bdI_N\right)$.

\item Include frequency and phase estimation errors: 
$\bdf(k-1)=\bdf(k-1)+\bm{\varepsilon}_f$ where 
$\bm{\varepsilon}_f\sim\N\left(\bdzero,\left(\sigma^m_f\right)^2\bdI_N\right)$, \\
$\bdtheta(k-1)=\bdtheta(k-1)+\bm{\varepsilon}_\theta$ where 
$\bm{\varepsilon}_\theta\sim\N\left(\bdzero,\left(\sigma^m_\theta\right)^2\bdI_N\right)$, 

\item Run the $k$-th iteration of the consensus algorithm as follow.
\begin{align}
\bdf(k)&=\bdW \bdf(k-1)\nonumber\\
\bdtheta(k)&=\bdW \bdtheta(k-1),\nonumber
\end{align}

\end{enumerate}
}
\KwOut{$\bdf(k)$, $\bdtheta(k)$}
\caption{DFPC Algorithm}
\end{algorithm}

\subsection{Steady-State Total Phase Error}\label{steady_state_section}


At the convergence of the proposed DFPC algorithm, the residual frequency and phase errors 
result from the frequency drifts and phase jitters in the oscillators, and due to the errors in 
estimating the frequencies and phases of their generated signals. Let each $k$-th iteration of the 
proposed algorithm be that of an update interval $T$. Now if the frequency and phase estimation is performed 
by observing the signals in this time duration, then in general increasing $T$ reduces the 
frequency and phase estimation errors but increases the errors due to the frequency drifts in the 
oscillators. 
In this subsection, we theoretically derive the residual steady-state total phase error for the DFPC algorithm 
as follow.

First lets assume that the frequency and phase estimation errors are zero, then the residual frequency error is 
caused by the frequency drift $\delta f_n$ of the oscillator at the $n$-th node. This drift 
is measured by the following standard deviation \cite{Hassna_TWC}
\begin{equation}
\sigma_f=f_c\sqrt{\frac{\beta_1}{T}+\beta_2 T},
\end{equation}
where the above equation combines the frequency deviation in the oscillator due to the white frequency noise which 
is measured by $\sigma_{wf}=f_c\sqrt{\frac{\beta_1}{T}}$ and the frequency deviation due to the frequency 
random walk which is quantified by $\sigma_{rw}=f_c\sqrt{\beta_2 T}$. 
The standard deviation of the resulting phase error due to the 
frequency deviation at time $T$ is given by
\begin{equation}
\sigma^f_\phi=2\pi\sigma_f T.
\end{equation}
Now, in practice, the frequency drift of the oscillator 
varies over time duration $T$ which causes the phase error $\delta \theta^f_n$ in the generated signal as derived 
in Section \ref{modeling_ftheta}. The standard deviation of this phase error is given by 
\begin{equation}
\sigma^p_\theta=\pi T \sigma_f.
\end{equation}
Furthermore, the phase noise in the oscillator induces a phase jitter in its generated signal which 
as discussed in Section \ref{modeling_ftheta} can be measured by the following standard deviation 
\begin{equation}
\sigma_\theta=\sqrt{2\times 10^{A/10}},
\end{equation} 
where as defined earlier 
$A$ is the integrated phase noise power computed from the phase noise profile of an oscillator \cite{Serge_Access_2021}. 

Next we derive the phase errors resulting from the frequency and phase estimation of the nodes output signals. 
Let the signal generated by the $n$-th node over time duration $T$ is given by 
\begin{equation}
x_n(t)=e^{j\left(2\pi f_nt+\theta_n\right)}+n_o(t),
\end{equation}
where $f_n$ and $\theta_n$ represent the frequency and phase of the signal respectively, 
and $n_o(t)$ is the transceiver noise which is modeled as a Gaussian process. The variation of the 
frequency estimation error is lower bounded by the following standard deviation 
\cite{Richards_radar} 
\begin{equation}\label{sigma_m_f}
\sigma^m_f\geq\sqrt{\frac{6}{(2\pi)^2 L^3 \text{SNR}}},
\end{equation}
in which $L$ denotes the number of samples collected over the time interval $T$, and SNR is the 
signal to noise ratio of the received signal. Similarly the variation in the phase estimation error is lower bounded by  
\begin{equation}\label{sigma_m_theta}
\sigma^m_\theta\geq \frac{2}{ L \text{SNR}}.
\end{equation}
Note that by comparing \eqref{sigma_m_f} and \eqref{sigma_m_theta}, it can be seen that the lower bound on the 
frequency estimation error ($\sigma^m_f$) is larger in value 
than the bound on the phase estimation error ($\sigma^m_\theta$) by 
$\sqrt{\frac{(3/2)\text{SNR}}{(2\pi)^2 L}}$. This implies that in general accurate frequency estimation is difficult to achieve 
from the observed signal samples compared to the phase estimation. However, to synchronize the nodes in a distributed phased array system, the collective phase error originated from 
the combined frequency and phase estimation errors is of importance to ensure that they system operates in a phase-coherent state.

Additionally, the nodes may enable a local broadcast to share 
their frequencies and phases with the other nodes for synchronization. 
For a node with $D$ number of connections on average and 
synchronized frequencies and phases with its neighbors, 
the estimation errors usually decrease with the increase in $D$ due to the increase in SNR. 
In this case, the lower bound for the frequency estimation is given by
\begin{equation}
\sigma^m_f\approx \sqrt{\frac{6 }{(2\pi)^2 D L^3 \text{SNR}}},
\end{equation}
and the 
bound for the phase estimation becomes
\begin{equation}
\sigma^m_\theta\approx\frac{2}{\sqrt{D} L \text{SNR}},
\end{equation}
where the approximation in the above equation is used because $D$ represents the average number of connections 
per node in a network. 
In general, with an increase in the number of connections per node, 
the resources required to support the wireless connections in a network 
also must increase proportionally. A solution to cope with the limited resources is to separate the node transmissions 
using either CDMA or TDMA. 
While this approach is deemed useful, one major consequence of 
using TDMA is that the observation time of the signals reduces which in turn reduces the number of samples collected 
for the frequency and phase estimation. For instance, 
with $D$ average number of connections per node, the collected number of samples $L$ reduces to $L/D$. Thus, when 
TDMA is used in the network, the lower bound for the frequency estimation is given by 
\begin{equation}\label{freq_est_TDMA}
\sigma^m_f\approx \sqrt{\frac{6 D^2}{(2\pi)^2 L^3 \text{SNR}}},
\end{equation}
which in terms of the phase error can be written as $\sigma^m_\phi=2\pi f_c\sigma^m_f T$. Similarly, the lower 
bound for the phase estimation becomes 
\begin{equation}\label{phase_est_TDMA}
\sigma^m_\theta\approx\frac{\sqrt{4D}}{L \text{SNR}}.
\end{equation}

Now after one iteration of the proposed DFPC algorithm, the total standard deviation 
due to frequency errors is $\sqrt{\sigma^2_f+\left(\sigma^m_f\right)^2}$ and the total standard 
deviation due to the phase errors is 
$\sqrt{\left(\sigma^p_\theta\right)^2+\left(\sigma^m_\theta\right)^2+\left(\sigma_\theta\right)^2}$. Thus to 
derive the total phase error after $I$ iterations, we model the frequency and phase update in the $k$-th 
iteration as
\begin{equation}
\bdz(k)=\bdW\left(\bdz(k-1)+\bde_k\right),
\end{equation}
in which $\bdz\in\{\bdf,\bdtheta\}$ and the error vector $\bde_k$ is distributed as 
$\N\left(\bdzero,\sigma^2_e\bdI_N\right)$. When $\bdz=\bdf$, the variance $\sigma^2_e$ is given by  
$\sigma^2_e=\sigma^2_f+\left(\sigma^m_f\right)^2$, 
and when $\bdz=\bdtheta$, the variance is 
$\sigma^2_e=\left(\sigma^p_\theta\right)^2+\left(\sigma^m_\theta\right)^2+\left(\sigma_\theta\right)^2$.
After $I$ number of iterations, the above update equation can be written as
\begin{align}\label{error_iter}
\bdz(I)&=\bdW\bdz(I-1)+\bdW\bde_I\nonumber\\
&=\bdW^2\bdz(I-2)+\bdW^2\bde_{I-1}+\bdW\bde_I\nonumber\\
&=\bdW^I\bdz(0)+\underbrace{\sum^{I-1}_{i=0}\bdW^{i+1}\bde_{I-i}}_{\text{total residual error}}.
\end{align}
Note that for a symmetric and a doubly stochastic matrix $\bdW$ as defined in \eqref{mix_W}, as 
$I\rightarrow\infty$ then $\bdW^I\rightarrow\frac{\bdone\bdone^T}{N}$ \cite{AC_2004}. 
This implies that 
$\bdW^I\bdz(0)$ in \eqref{error_iter} converges to the average of the initial values, i.e., as $I\rightarrow \infty$, 
$\bdW^I\bdz(0){\rightarrow} \frac{\bdone\bdone^T\bdz(0)}{N}=\bdone\left(\frac{1}{N}\sum^N_{n=1}z_n(0)\right)$ 
in which $z_n(0)$ is the $n$-th element of the vector $\bdz(0)$. Similarly, for a large $I$, the dominant term 
in the total residual error in \eqref{error_iter} is $\bdW^I\bde_1$ that converges to the vector 
$\frac{\bdone\bdone^T\bde_1}{N}$ where the elements in this vector represent the average of the elements 
in $\bde_1$ (the error vector in the first iteration). Thus to 
quantify the total residual error after a large number of iterations, 
we find the covariance of the following term
\begin{align}\label{cov_e}
&\sum^{I-1}_{i=0}\bdW^{i+1}\bde_{I-i}-\frac{1}{N}\sum^{I-1}_{i=0}\bdone\bdone^T\bde_{I-i}\nonumber\\
&=\sum^I_{m=1}\left(\bdW-\frac{1}{N}\bdone\bdone^T\right)^m\bde_{I-m+1},
\end{align}
where to get the equality in the above equation, we performed a change of summation variable by choosing 
$m=i+1$ and used the identity in Eqn. (11) in \cite{AC_2004}. The covariance matrix of the error term in \eqref{cov_e} 
is given by
\begin{align}
\sigma^2_e\sum^I_{m=1}\left(\bdW-\frac{1}{N}\bdone\bdone^T\right)^{2m}.
\end{align}

Now for $\lambda_2$ defined as the second eigenvalue modulus of the mixing matrix $\bdW$, it can be easily shown that 
$\lambda^{2m}_2\bdI_N-\left(\bdW-\frac{1}{N}\bdone\bdone^T\right)^{2m}$ is a positive semi-definite matrix. 
Thus, after $I$ iterations the standard deviation of the total residual error can be given as 
\begin{align}
\sigma_{e,\text{residual}}=\sqrt{\sigma^2_e\sum^I_{m=1}\lambda^{2m}_2}.
\end{align}
Note that when the network is sparsely connected then $\lambda_2$ is close to $1$ and $\sum^I_{m=1}\lambda^{2m}_2\gg 1$ 
which implies a higher residual error. At a specific value of network connectivity, we have $\sum^I_{m=1}\lambda^{2m}_2=1$ 
which results in $\sigma_{e,\text{residual}}=\sigma_e$ and the standard deviation of the total phase error is given 
by 
\begin{equation}\label{sigma_total}
\sigma_{\phi,\text{total}}=\sqrt{\left(\sigma^f_\phi\right)^2+\left(\sigma^m_\phi\right)^2+\left(\sigma^p_\theta\right)^2+\left(\sigma^m_\theta\right)^2+\left(\sigma_\theta\right)^2}.
\end{equation}
However, as the connectivity in the network increases, the second eigenvalue modulus $\lambda_2$ approaches $0$, and thus the standard deviation of the total phase error decreases below $\sigma_{\phi,\text{total}}$. This change in the total 
phase error with the change in network connectivity is demonstrated through simulations in Section \ref{residual_phase_section}. 
\begin{figure*}[t!]
    \begin{minipage}{0.34\textwidth}
        \centering
\includegraphics[width=1.0\textwidth,height=0.65\textwidth]{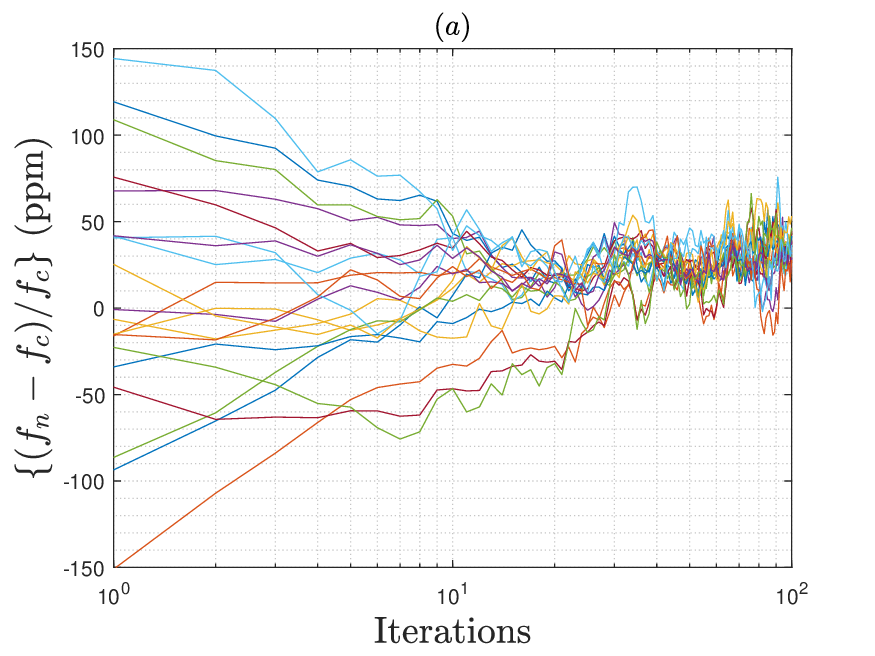}
    \end{minipage}
    \begin{minipage}{0.34\textwidth}
        \centering
\includegraphics[width=1.0\textwidth,height=0.65\textwidth]{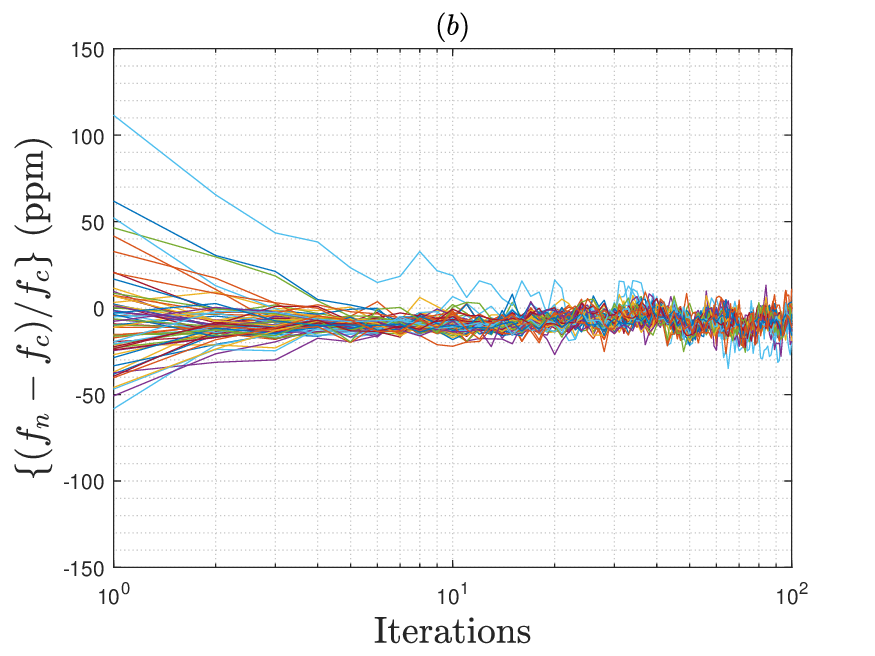}
		\end{minipage}
    \begin{minipage}{0.34\textwidth}
        \centering
\includegraphics[width=1.0\textwidth,height=0.65\textwidth]{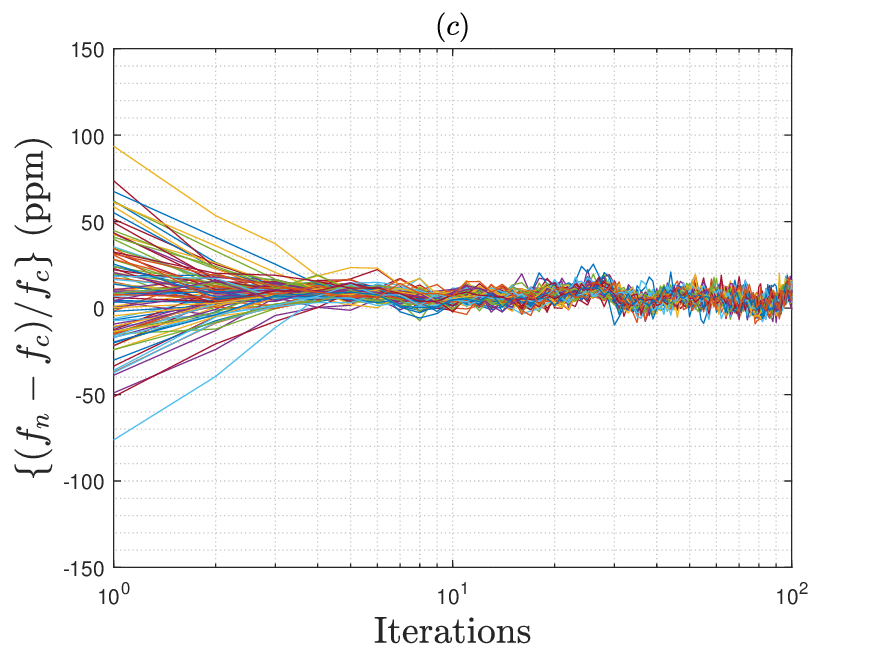}
		\end{minipage}
		\caption{Frequency errors for all $N$ nodes in the network vs. iterations of DFPC for $c=0.2$, $\text{SNR}=0$ dB, 
		and $T=0.1$ ms when $(a)$ $N=20$, $(b)$ $N=65$, and $(c)$ $N=100$.}
		\label{fig:freqs}
\end{figure*}
\begin{figure*}[t!]
    \begin{minipage}{0.34\textwidth}
        \centering
\includegraphics[width=1.0\textwidth,height=0.65\textwidth]{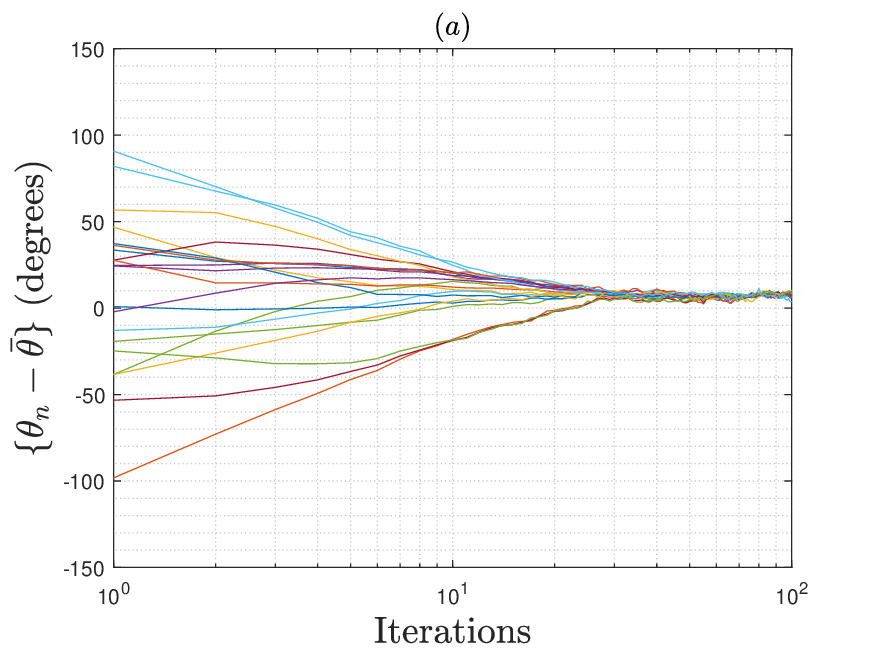}
    \end{minipage}
    \begin{minipage}{0.34\textwidth}
        \centering
\includegraphics[width=1.0\textwidth,height=0.65\textwidth]{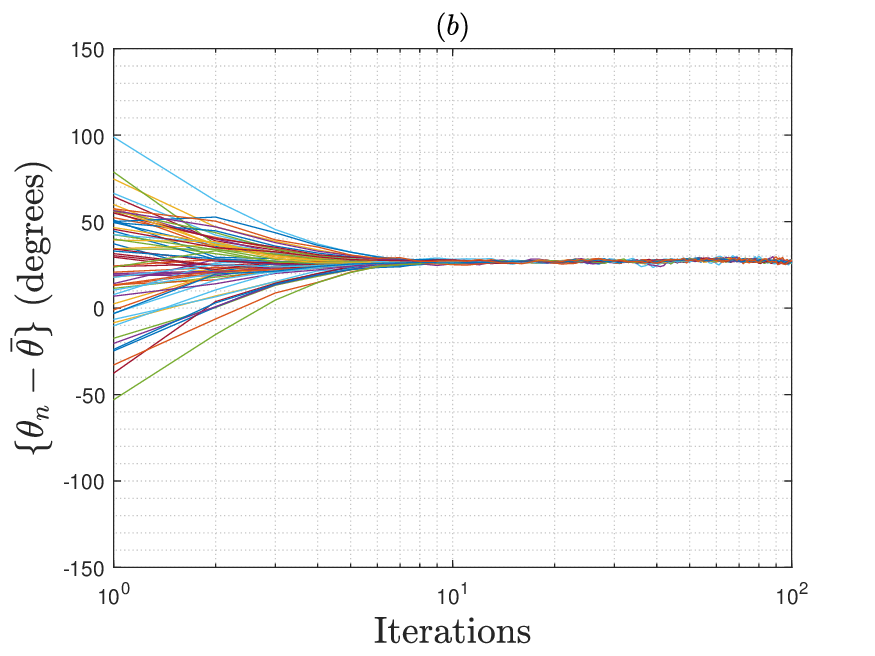}
		\end{minipage}
    \begin{minipage}{0.34\textwidth}
        \centering
\includegraphics[width=1.0\textwidth,height=0.65\textwidth]{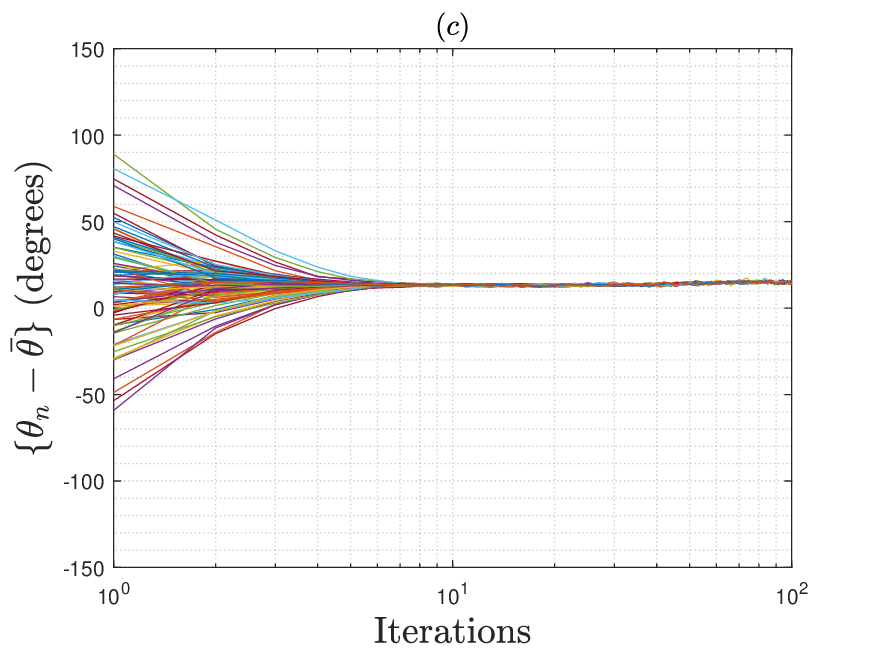}
		\end{minipage}

		\caption{Phase errors for all $N$ nodes in the network vs. iterations of DFPC for $c=0.2$, $\text{SNR}=0$ dB, 
		and $T=0.1$ ms when $(a)$ $N=20$, $(b)$ $N=65$, and $(c)$ $N=100$.}

		\label{fig:phases}
\end{figure*}

\subsection{Simulation Results}\label{DFPC_sim}
In this subsection, we analyze the synchronization performance of the proposed DFPC algorithm 
through simulations. 
We consider the case when a network of $N$ nodes is 
randomly generated with a known connectivity $c$. The parameter $c$ is defined 
as the ratio of the number of active edges in the network to the number of all possible edges 
given by $N(N-1)/2$. Thus the connectivity $c\in[0,1]$ where 
a smaller value of $c$ implies a sparsely connected network and a larger value of $c$ 
implies a densely connected network.
Throughout the simulation results included in this paper 
the initial frequencies of all the nodes are sampled from the normal distribution centered on $f_c=1 \text{ GHz}$ 
and the sampling frequency is chosen as $f_s=10$ MHz. 
Since for coherent distributed arrays, the update interval $T$ 
is usually on the order of ms to sec, we choose $T=0.1$ ms for 
the results in this paper unless stated otherwise. 

In Figs. \ref{fig:freqs} and \ref{fig:phases} we show the frequency and phase errors for different 
number of nodes $N$ in the network vs. the number of iterations from a single run of the DFPC algorithm. 
The network is assumed to be sparsely connected with $c=0.2$ and the update interval is set as $T=0.1$ ms. 
The $\text{SNR}=0$ dB is used to compute the frequency and phase estimation errors from 
\eqref{f_est_eqn} and \eqref{theta_est_eqn}. It is observed that as the number of iterations of the 
DFPC algorithm increases the frequencies and phases of all the nodes in all simulated cases converge to an average of their 
initial values with some residual errors. These residual frequency and phase errors upon convergence result from 
the frequency drifts and phase jitters in the oscillators, and the frequency and phase estimation errors at the nodes 
as derived in the previous section and 
integrated in the synchronization process to imitate a more practical scenario. However, it will be shown later herein 
that the variation in the residual total phase error is well below 
the $18^\deg$ threshold which ensures high coherence at the destination \cite{OCDA_2017}.
From these figures, it is also observed that the convergence speed of the DFPC algorithm is faster 
for larger number of nodes ($N=100$) vs. smaller 
number of nodes ($N=20$). This is because as $N$ increases for a given connectivity $c$, there are more connections 
per node on average given by $D=c(N-1)$, 
and thus each node computes a better local average in the network resulting in a faster convergence. 
Therefore, for a given connectivity $c$, larger number of nodes in a network makes the system more probable 
to converge to the averages of the initial frequency and phase distributions which is as also visible from these figures.
Note that while we plot the frequency and phase errors relative to the average values, convergence 
to a specific frequency and phase is not a compulsory condition 
for a coherent operation in distributed arrays. In fact, 
consensus at any frequency and phase will support the coherent operation providing the residual errors are 
small as discussed earlier.

\begin{figure*}[tp]
		
    \begin{minipage}{0.48\textwidth}
        \centering
\includegraphics[width=0.99\textwidth]{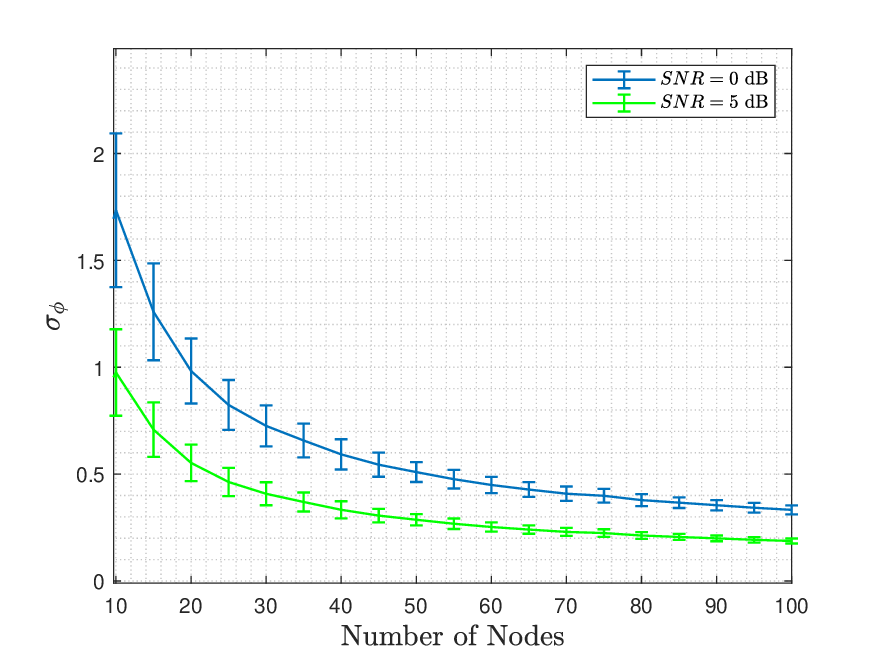}
	\caption{Standard deviation of the total phase error of DFPC vs. 
	the number of nodes $N$ in the network for different $\text{SNRs}$ when $c=0.2$ and $T=0.1$ ms.}
	\label{fig:sigma_vs_N_DFPC}
		\end{minipage}\hspace{.025\linewidth}
    \begin{minipage}{0.48\textwidth}
        \centering
     	\includegraphics[width=0.99\textwidth]{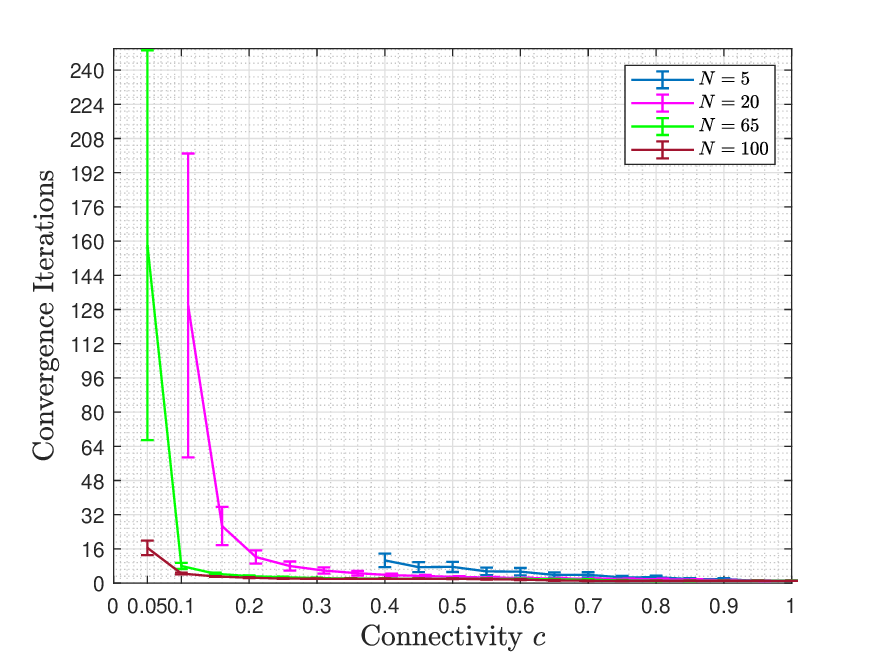}
	\caption{Covergence iterations of DFPC vs. connectivity $c$ in the network with different 
	number of nodes when $\text{SNR}=0$ dB and $T=0.1$ ms.}
		\label{fig:iter_vs_c_DFPC}
					\vspace{.2cm}
    \end{minipage}
\end{figure*}

Fig. \ref{fig:sigma_vs_N_DFPC} shows the standard deviation of the total phase errors 
($\delta \phi_n$, for $n=1,2,\ldots,N$ as defined in \eqref{Sig_model} in Section \ref{modeling_section}) of the 
DFPC algorithm 
for varying number of nodes $N$ in the network when two different $\text{SNRs}$ are assumed. The connectivity 
in the network is set to $c=0.2$ and the update interval is $T=0.1$ ms. 
To generate this figure, we collected the final converged standard deviation of the total phase errors 
from $10^3$ independent trials and then computed the average value and the standard 
deviation of the samples. In this figure, the length of the bar determines the value of the 
standard deviation whereas its center gives the average value at each point. 
It is observed that for each $\text{SNR}$ value as the number of nodes $N$ in the network increases the 
total phase error and its variation decreases. As mentioned before, this is due to the increase 
in the value of $D$ which aids in computing more accurate frequency and phase averages per node. 
However, for larger values of $N$ the total phase errors reaches a noise floor resulting from the 
frequency and phase offset errors and the estimation errors. Increase in $\text{SNR}$ from $0$ dB to $5$ dB shows 
improvement in performance because the residual phase errors due to the frequency and phase estimation decreases. 
Note that when there are no frequency and phase offset errors, the total residual phase error is $0^\deg$ 
upon convergence.

In Fig. \ref{fig:iter_vs_c_DFPC} we compare the convergence speed of DFPC for different number of nodes 
vs. the connectivity $c$ of the network. To generate this figure, we declare the convergence of DFPC 
when the standard deviation of the total phase error over the iterations is less than {$\eta=1^\deg$ threshold 
which is well below the $18^\deg$ threshold as needed to achieve at least $90\%$ coherent gain at the destination \cite{OCDA_2017}.}
In practice, a particular value of the threshold may be application dependent. 
In this figure, the number of iterations needed for convergence 
are sampled from $10^3$ independent trials to plot the average value 
and the standard deviation of the samples. Note that for $N=5$ and $20$ nodes the lowest possible 
$c$ values that ensure a connected network are $0.4$ and $0.11$, respectively, and that is why the other values of 
$c$ are not simulated for these many nodes in the network. 
It is observed that for all $N$ values, the convergence speed of DFPC 
improves with the increase in the connectivity $c$ of the network as expected. Furthermore, for each connectivity 
level $c$ having more nodes in the network result in faster convergence. 
In particular, DFPC for $N=100$ converges 
faster for a moderately connected networks with $c\in[0.05, 0.8]$. For example, at connectivity $c=0.05$, 
DFPC with $N=100$ nodes requires $16$ iterations on average for the convergence whereas when 
$N=65$ the convergence happens in $158$ iterations. 
Besides the change in $D$ value with the increase of $c$ or $N$, 
this improvement in the convergence performance of DFPC 
with varying $c$ or $N$ can also be attributed to 
the modulus of the second largest eigenvalue $\lambda_2$ of the mixing matrix $\bdW$. 
It is shown in Section \ref{steady_state_section} that $\lambda_2$ controls the total phase error of DFPC. 
Specifically, $\lambda_2$ is smaller for denser networks and larger for sparser networks, and thus 
the algorithm converges faster for the former networks than the latter ones. 

\section{Kalman Filter based Decentralized Frequency and Phase Synchronization}\label{KF_freq_phase_sync}

In the previous section, we observed that the frequency drifts and phase jitters as well as the 
frequency and phase estimation errors at the nodes introduce the total phase 
error upon convergence of the DFPC algorithm that deteriorates the synchronization between the nodes. 
{An increase in $\text{SNR}$ reduces the 
phase error of DFPC, but given the multipath fading channels between 
the nodes, an improvement in $\text{SNR}$ can be achieved by increasing the signal power, which may 
be limited by the hardware constraints.} 
The Kalman filter is a popular method used for computing optimal the minimum mean square error (MMSE) 
estimates of the unknown quantities if their state transitioning model follows a first-order Markov process 
and the observations are a linear function of these quantities \cite{Anderson1979, Claser_2021}. 
Recently, KF has been used for the time synchronization between the nodes in 
\cite{KF-TS_2015, KF-TS_2019, KF-TS_2021} where the oscillators time drifting models are exploited. 
Thus, in this section, to reduce the residual total phase error {at the faster update intervals} 
we propose a Kalman filtering based 
decentralized consensus algorithm for the frequency and phase synchronization between the nodes in a distributed array. 
The proposed algorithm is referred to as KF-DFPC and it integrates Kalman filtering with 
the DFPC algorithm described earlier to improve synchronization between the nodes. 
{Alternatively, the model-free adaptive filtering algorithms such as the 
diffusion least mean squares algorithm \cite{LMS_2010, Claser_2021}, and the KF-based 
distributed state estimation algorithms in \cite{Saber_2009, Xin_2022, Sayed_2010} can be used; 
however, it is shown through the simulation results later in this section that 
our proposed KF-DFPC algorithm outperforms and converges faster than these 
contemporary algorithms at lower SNRs and for the sparsely 
connected arrays, whereas for the large densely 
connected array, the use of KF does not increase the computational complexity of DFPC.}

In order to use Kalman filter for the frequency and phase estimation at the nodes, 
we need to define their state transitioning model and the observation model as per 
its framework \cite{Anderson1979}. Let at the $k$-th time instant, 
the unknown frequency and phase of the $n$-th node can be defined as a state vector 
$\bdx_n(k)=[f_n(k),\theta_n(k)]^T$. Using the frequency and phase offset models discussed 
in Section \ref{modeling_ftheta}, we can write the state transition model for the $n$-th 
node as 
\begin{equation}\label{state_eqn}
\bdx_n (k)=\bdx_n(k-1)+\bdu_n,
\end{equation}
where $\bdu_n\triangleq\left[\delta f_n,\delta \theta^f_n+\delta \theta_n\right]^T$ and 
$\bdu_n\sim\N(\bdzero,\bdQ)$ 
in which the correlation matrix $\bdQ$ is given by 
\begin{equation}
\bdQ=\Exp[\bdu_n\bdu^T_n]=
\left[
\begin{matrix}
\sigma^2_f && -\pi T\sigma^2_f\\
-\pi T\sigma^2_f && \pi^2T^2\sigma^2_f+\sigma^2_\theta
\end{matrix}
\right].
\end{equation}
\begin{figure*}[b]
\hrulefill
\normalsize
\setcounter{mytempeqncnt}{\value{equation}}
\setcounter{equation}{34}
\begin{align}\label{time_upd_m}
\bdm_{n,k}(k)
&=\bdm_{n,k-1}(k)+\bdV_{n,k-1}(k)\left(\bdV_{n,k-1}(k)+\bdSigma\right)^{-1}\left(\bdy_n(k)-\bdm_{n,k-1}(k)\right),\\
\bdV_{n,k}(k)&=\bdV_{n,k-1}(k)-\bdV_{n,k-1}(k)\left(\bdV_{n,k-1}(k)+\bdSigma\right)^{-1}\bdV_{n,k-1}(k),\label{time_upd_V}
\end{align}
\setcounter{equation}{\value{mytempeqncnt}}
\end{figure*}

As seen from Equation \eqref{state_eqn}, the frequency and phase of the oscillator in each node's transceiver 
are influenced by random offsets, so their instantaneous values are unknown to the nodes and must be 
estimated for synchronization. Let the vector $\bdy_n(k)=\left[\hat{f}_n(k),\hat{\theta}_n(k)\right]^T$ 
define the observation vector that combines the frequency and phase estimates of the $n$-th node's signal 
obtained from an estimator as discussed in Section \ref{modeling_ftheta}, 
then the observation model for the $n$-th node can be written as
\begin{equation}\label{obs_eqn}
\bdy_n(k)=\bdx_n(k)+\bdv_n,
\end{equation}
where $\bdv_n\triangleq\left[\varepsilon_f,\varepsilon_\theta\right]^T$ in which $\varepsilon_f$ and 
$\varepsilon_\theta$ represent the frequency and phase estimation errors respectively. 
Assuming that the frequency and phase estimation errors are independent, 
we can model the estimation error vector as $\bdv_n\sim\N(\bdzero,\bdSigma)$ where the correlation matrix 
$\bdSigma$ is given by 
\begin{equation}
\bdSigma=\Exp[\bdv_n\bdv^T_n]=\left[
\begin{matrix}
\left(\sigma^m_f\right)^2 && 0\\
0 && \left(\sigma^m_\theta\right)^2
\end{matrix}
\right].
\end{equation} 
The parameters $\sigma^m_f$ and $\sigma^m_\theta$ represent the standard deviation of the frequency and phase 
estimation errors respectively. For the purpose of illustrating the synchronization performance of the proposed  
algorithm, herein 
both $\sigma^m_f$ and $\sigma^m_\theta$ are set equal to the CRLBs given in \eqref{f_est_eqn} and 
\eqref{theta_est_eqn} respectively. Note 
that the CRLB can be achieved by an unbiased and efficient estimator that estimates the frequency and phase by collecting 
a large number of samples over the observation window at the cost of latency. 
\begin{algorithm}\label{algo_2}
  \footnotesize
\DontPrintSemicolon
\SetKwInput{KwPara}{Input}
\SetKwFor{ForEach}{for each}{}{end}
\KwPara{$k=0$, $\bdW$, $\bdf(0)$, $\bdtheta(0)$, and $\bdm_{n,0}(0)$, $\bdV_{n,0}(0)$, for $n=1,2,\ldots,N$.}
\tcc{KF-DFPC run}
\While{convergence criterion is not met} 
{
	$k=k+1$
\begin{enumerate}
\item Define $\bdf(k)=\bdf(k-1)+\bddelta \bdf$ where frequency drifts \\are modeled as 
$\delta \bdf\sim \mathcal{N}\left(\bdzero,\sigma^2_f\bdI_N\right)$.

\item Set $\bdtheta(k)=\bdtheta(k-1)+\bddelta \bdtheta^f+\bddelta \bdtheta$ 
where phase errors 
$\bddelta \bdtheta^f\triangleq\left[\delta \theta^f_1,\delta \theta^f_2, \ldots, \delta \theta^f_N\right]^T$ 
are computed from \eqref{theta_f_n}\\ 
and phase jitters are modeled as 
$\bddelta \bdtheta\sim \N \left(\bdzero,\sigma^2_\theta \bdI_N\right)$.

\item Include frequency and phase estimation errors: \\
$\hat{\bdf}(k)=\bdf(k)+\bm{\varepsilon}_f$ where 
$\bm{\varepsilon}_f\sim\N\left(\bdzero,\left(\sigma^m_f\right)^2\bdI_N\right)$, \\
$\hat{\bdtheta}(k)=\bdtheta(k)+\bm{\varepsilon}_\theta$ where 
$\bm{\varepsilon}_\theta\sim\N\left(\bdzero,\left(\sigma^m_\theta\right)^2\bdI_N\right)$.\\  
\ForEach{$n=\{1,2,\ldots,N\}$}{
\begin{enumerate}[leftmargin=-0.1cm]
\item Define the observation vector $\bdy_n=\left[\hat{f}_n(k),\hat{\theta}_n(k)\right]^T$.\\
\eIf{$k=1$}
{
\begin{enumerate}[leftmargin=-0.1cm]
\item Run the prediction update of KF by computing \\ 
$\bdm_{n,k-1}(k)$ and $\bdV_{n,k-1}(k)$ from \eqref{pred_upd}.
\end{enumerate}
}
{
\begin{enumerate}[leftmargin=-0.1cm]
\item Set $\bdm_{n,k-1}(k-1)=\left[f_n(k-1),\theta_n(k-1)\right]^T$\\ and compute 
$\bdV_{n,k-1}(k-1)$ from \eqref{init_KF_V}.
\item Run the prediction update of KF by finding \\
$\bdm_{n,k-1}(k)$ and $\bdV_{n,k-1}(k)$ using \eqref{pred_upd}.
\end{enumerate}

}
\item Run the time update step of KF by computing \\ 
$\bdm_{n,k}(k)$ and $\bdV_{n,k}(k)$ using \eqref{time_upd_m} and \eqref{time_upd_V}.
\end{enumerate}
}
\item For $\bdm_{n,k}(k)\triangleq\left[m^f_{n,k}(k),m^\theta_{n,k}(k)\right]^T$ $\forall$ $n=1,2,\ldots,N$, \\ 
define 
$\bdm^f_k(k)=\left[m^f_{1,k}, m^f_{2,k},\ldots,m^f_{N,k}\right]^T$ and
$\bdm^\theta_k(k)=\left[m^\theta_{1,k}, m^\theta_{2,k},\ldots,m^\theta_{N,k}\right]^T$.
\item Update the frequencies and phases of all the nodes by using
\begin{align*}
\bdf(k)&=\bdW\bdm^f_k(k)\\
\bdtheta(k)&=\bdW\bdm^\theta_k(k),
\end{align*} 
\end{enumerate}

}
\KwOut{$\bdf(k)$, $\bdtheta(k)$}
\caption{KF-DFPC Algorithm}
\end{algorithm}

The KF algorithm is an iterative algorithm that estimates the unknown state vector 
in each iteration by computing the prediction update step and the time update step. 
In the prediction update step of the $k$-th iteration, it predicts the posterior distribution 
on the state vector $\bdx_n(k)$ given the observations up to the time instant $k-1$. 
The posterior distribution is computed as
\begin{align}\label{pred_post}
&p\left(\bdx_n(k)\mid \bdy^{1:k-1}_n\right)\nonumber\\
&=\int \left[p\left(\bdx_n(k)|\bdx_n(k-1)\right)p(\bdx_n(k-1)|\bdy^{1:k-1}_n)\right]d\bdx_n(k-1)\nonumber\\
&\propto \N\left(\bdm_{n,k-1}(k),\bdV_{n,k-1}(k)\right),
\end{align}
where in \eqref{pred_post} the state transition distribution is 
$p\left(\bdx_n(k)|\bdx_n(k-1)\right)=\N\left(\bdx_n(k-1),\bdQ\right)$, i.e., it is normally distributed 
as per the distribution of the offset vector in \eqref{state_eqn}. Assuming that the conditional distribution 
$p(\bdx_n(k-1)|\bdy^{1:k-1}_n)=\N\left(\bdm_{n,k-1}(k-1),\bdV_{n,k-1}(k-1)\right)$, we solve the 
convolution of the two normal distributions in \eqref{pred_post} to get the mean and the covariance of the resulting 
posterior distribution as
\begin{align}\label{pred_upd}
\bdm_{n,k-1}(k)&=\bdm_{n,k-1}(k-1)\nonumber\\
\bdV_{n,k-1}(k)&=\bdV_{n,k-1}(k-1)+\bdQ.
\end{align}
Thus, the prediction update step uses the state vector estimate from the previous time instant to 
produce the estimate of the state vector at the current time instant. 
In the time update step of the $k$-th iteration, the KF algorithm combines the observation 
from the current time instant to refine this prediction of the state vector and 
hence obtain a more accurate state 
vector estimate. To this end, it computes the following posterior distribution
\begin{align}\label{time_upd_post}
p\left(\bdx_n(k)\mid \bdy^{1:k}_n\right)
&\propto p\left(\bdx_n(k)|\bdy^{1:k-1}_n\right)p(\bdy_n(k)|\bdx_n(k))\nonumber\\
&\propto \N\left(\bdm_{n,k}(k),\bdV_{n,k}(k)\right),
\end{align}
where in \eqref{time_upd_post} the likelihood function $p\left(\bdy_n(k)|\bdx_n(k)\right)$ is normally distributed as 
$\N(\bdx_n(k),\bdSigma)$ which is obtained by shifting the distribution of the estimation error vector in 
\eqref{obs_eqn} by $\bdx_n(k)$. Inserting \eqref{pred_post} in \eqref{time_upd_post}, the mean and the covariance of the 
resulting posterior distribution are computed in \eqref{time_upd_m} and \eqref{time_upd_V}, respectively. 
This completes the derivation of the Kalman filtering algorithm. 
\begin{figure*}[tp]
		
    \begin{minipage}{0.48\textwidth}
        \centering
\includegraphics[width=0.99\textwidth]{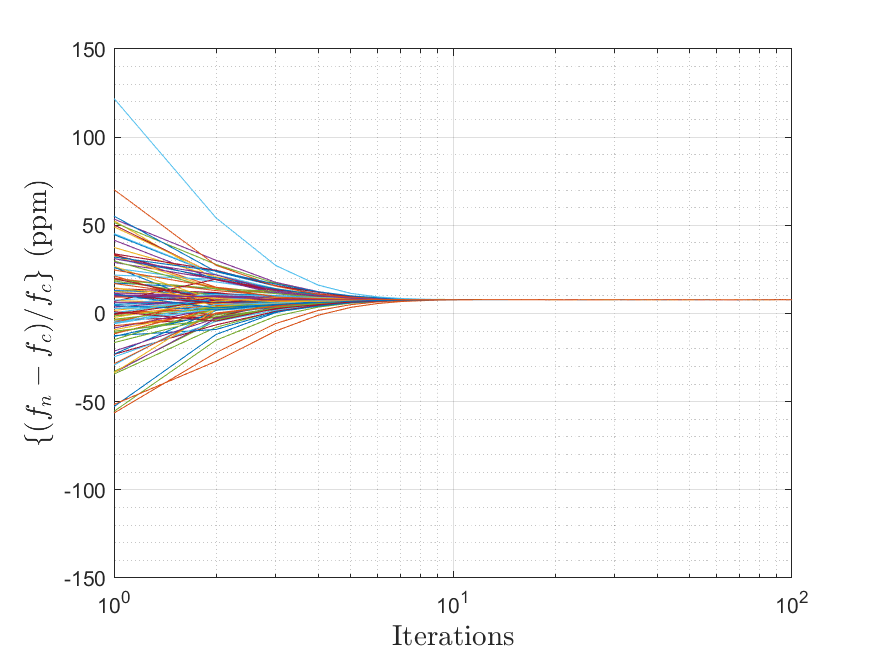}
	\caption{Frequency errors for $N=100$ nodes in the network  vs. 
	the iterations of KF-DFPC for $c=0.2$, $\text{SNR}=0$ dB, and $T=0.1$ ms.}
	\label{fig:freq_vs_iter_KFDFPC}
		\end{minipage}\hspace{.025\linewidth}
    \begin{minipage}{0.48\textwidth}
        \centering
     	\includegraphics[width=0.99\textwidth]{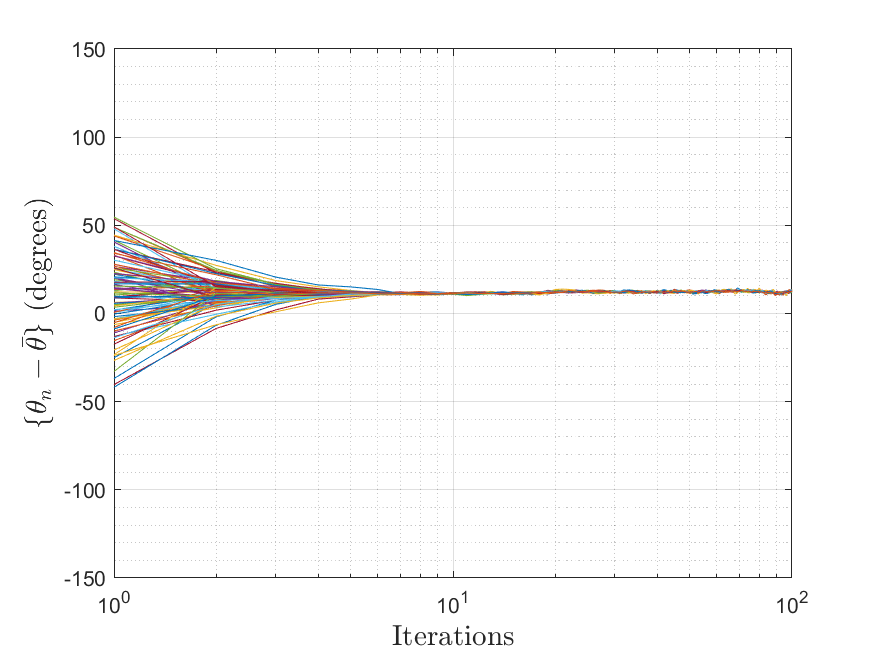}
	\caption{Phase errors for $N=100$ nodes in the network  vs. 
	the iterations of KF-DFPC for $c=0.2$, $\text{SNR}=0$ dB, and $T=0.1$ ms.}
		\label{fig:phase_vs_iter_KFDFPC}
					\vspace{.2cm}
    \end{minipage}
\end{figure*}
\begin{figure*}[tp]
		
    \begin{minipage}{0.48\textwidth}
        \centering
\includegraphics[width=0.99\textwidth]{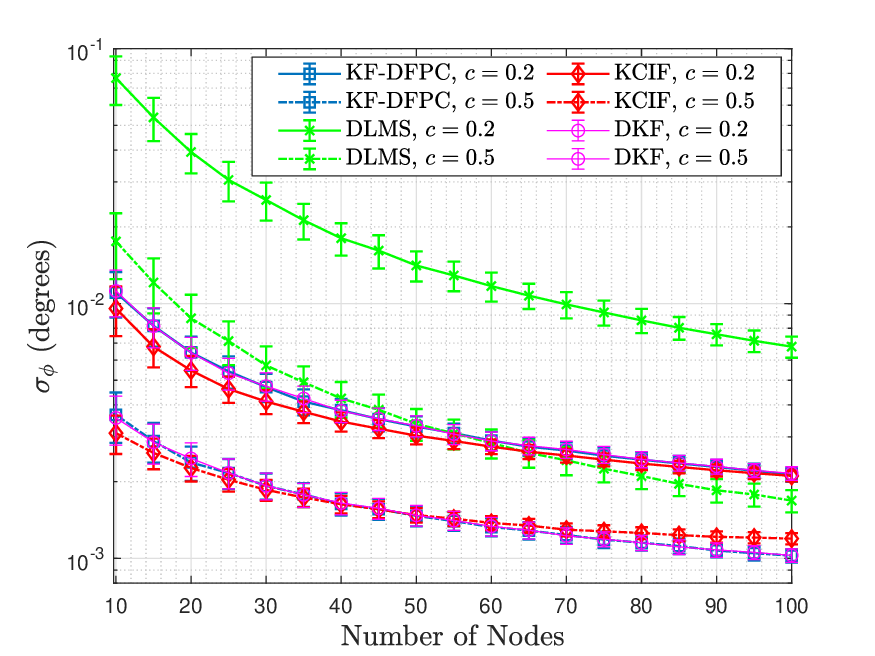}
	\caption{Standard deviation of the total phase error of KF-DFPC, DLMS, KCIF, and DKF vs. 
	the number of nodes $N$ in the network for different connectivity $c$ when $\text{SNR}=0$ dB and $T=0.1$ ms.}
	\label{fig:sigma_vs_N_KFDFPC}
		\end{minipage}\hspace{.025\linewidth}
    \begin{minipage}{0.48\textwidth}
        \centering
     	\includegraphics[width=0.99\textwidth]{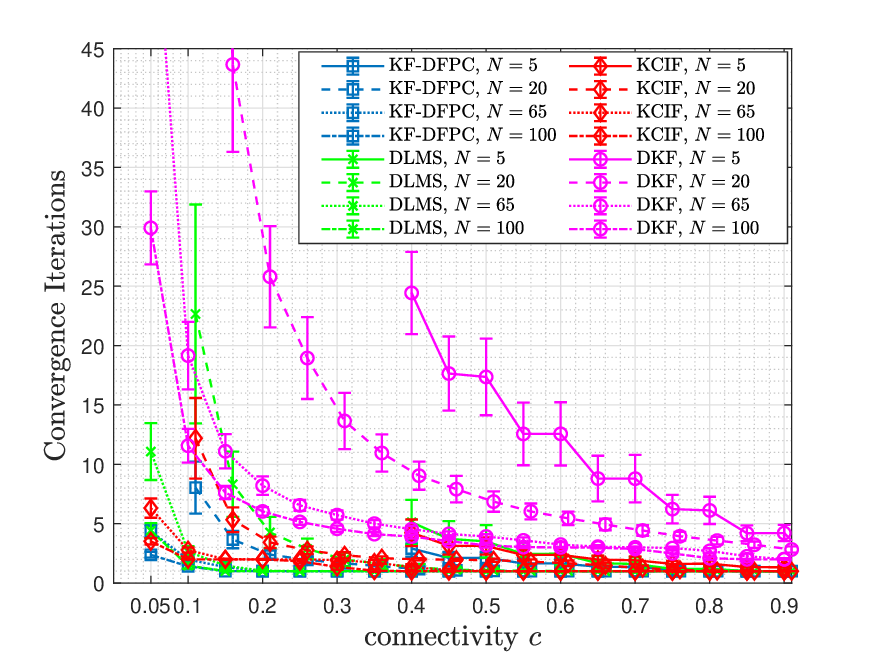}
	\caption{Convergence iterations of KF-DFPC, DLMS, KCIF, and DKF vs. connectivity $c$ in the network with different 
	number of nodes when $\text{SNR}=0$ dB and $T=0.1$ ms.}
		\label{fig:iter_vs_c_KFDFPC}
					\vspace{.2cm}
    \end{minipage}
\end{figure*}

Next we describe the proposed KF-DFPC Algorithm \ref{algo_2} which requires initialization of the KF 
algorithm in its each iteration. This initialization is explained as follow.  
To begin, in the $k=1$ iteration of the KF-DFPC algorithm, the prediction update Eqn. \eqref{pred_upd} of the Kalman 
filter at the $n$-th node can be initialized with the mean and variance of each node's 
initial frequency and phase distribution, as stated in Section \ref{DFPC_section}, as follows 
\addtocounter{equation}{2}
\begin{align}
\bdm_{n,0}(0)&=[f_c, \pi]^T\nonumber\\
\bdV_{n,0}(0)&=\left[\begin{matrix}\sigma^2&&0\\0&&4\pi^2/12\end{matrix}\right].
\end{align}
Now in the $k$-th iteration of KF-DFPC algorithm, let the state vector estimate of the $n$-th 
node after the time update step in \eqref{time_upd_m} can be written as 
$\bdm_{n,k}(k)=\left[m^f_{n,k}(k),m^\theta_{n,k}(k)\right]^T$ for all $n=1,2,\ldots,N$. By separating the 
frequency and phase estimates from all the nodes into vectors 
$\bdm^f_k(k)\triangleq\left[m^f_{1,k}, m^f_{2,k},\ldots,m^f_{N,k}\right]^T$ and
$\bdm^\theta_k(k)\triangleq\left[m^\theta_{1,k}, m^\theta_{2,k},\ldots,m^\theta_{N,k}\right]^T$, the KF-DFPC 
algorithm updates the frequencies and phases of all the nodes by using
\begin{align}\label{KF_DFPC_upd}
\bdf(k)&=\bdW\bdm^f_k(k)\nonumber\\
\bdtheta(k)&=\bdW\bdm^\theta_k(k),
\end{align}
where the decentralized mixing matrix $\bdW$ is defined in \eqref{mix_W}. 
This update implies that in the $k>1$ iterations of the KF-DFPC algorithm, the initialization of Kalman filter 
at the $n$-th node must reflect the linear frequency and phase transformation 
in \eqref{KF_DFPC_upd}. Since the posterior distribution on the state vector of the $n$-th node 
is a normal distribution, the mean and covariance after a linear transformation can be easily computed \cite{gubner_2006}. 
Thus the mean vector used for initializing \eqref{pred_upd} 
after the linear transformation in \eqref{KF_DFPC_upd} and notation adjustment is 
given by $\bdm_{n,k-1}(k-1)=\left[f_n(k-1),\theta_n(k-1)\right]^T$ where $f_n(k-1)$ and $\theta_n(k-1)$ denote 
the $n$-th element of $\bdf(k-1)$ and $\bdtheta(k-1)$ vectors, respectively. Similarly, in the $k>1$ iterations, the 
covariance matrix $\bdV_{n,k-1}(k-1)$ in \eqref{pred_upd} after the linear transformation in \eqref{KF_DFPC_upd} 
and notation adjustment can be given as 
\begin{align}\label{init_KF_V}
&\bdV_{n,k-1}(k-1)\nonumber\\
&=\left[
\begin{matrix}
\bdw^T_n\bdV^f_{k-1}(k-1)\bdw_n & \bdw^T_n\bdV^{f\theta}_{k-1}(k-1)\bdw_n\\
\bdw^T_n\bdV^{f\theta}_{k-1}(k-1)\bdw_n& \bdw^T_n\bdV^\theta_{k-1}(k-1)\bdw_n
\end{matrix}\right],
\end{align}
where the column vector $\bdw_n$ represents the $n$-th row of the mixing matrix $\bdW$. 
The diagonal matrix $\bdV^f_{k-1}(k-1)=\diag\{v^{1,1}_{1,k-1}(k-1),v^{1,1}_{2,k-1}(k-1),\ldots,v^{1,1}_{N,k-1}(k-1)\}$ 
in which each $v^{1,1}_{n,k-1}(k-1)$ for $n=1,2,\ldots, N$ 
is the $(1,1)$-th indexed element of the covariance matrix obtained 
from \eqref{time_upd_V} in the $(k-1)$-st iteration of KF-DFPC. 
Similarly, $\bdV^\theta_{k-1}(k-1)=\diag\{v^{2,2}_{1,k-1}(k-1),v^{2,2}_{2,k-1}(k-1),\ldots,v^{2,2}_{N,k-1}(k-1)\}$ 
where each $v^{2,2}_{n,k-1}(k-1)$ represents the $(2,2)$-th indexed element of the covariance matrix in 
\eqref{time_upd_V}, and 
the matrix $\bdV^{f\theta}_{k-1}(k-1)=\diag\{v^{1,2}_{1,k-1}(k-1),v^{1,2}_{2,k-1}(k-1),\ldots,v^{1,2}_{N,k-1}(k-1)\}$ 
in which each $v^{1,2}_{n,k-1}(k-1)$ denotes the $(1,2)$-th indexed element of the covariance matrix computed from 
\eqref{time_upd_V} in the $(k-1)$-st iteration of the KF-DFPC algorithm. 
This completes the derivation of the KF-DFPC algorithm which is described in detail in Algorithm \ref{algo_2}. 

{Note that for the notational convenience, 
we have written \eqref{init_KF_V} using the diagonal matrices $\bdV^f_{k-1}(k-1)$, 
$\bdV^{f\theta}_{k-1}(k-1)$, and $\bdV^\theta_{k-1}(k-1)$ for the entire array; however, since most of the elements in the 
weighting vector $\bdw_n$ may be zero as per the node's connectivity, only the diagonal elements corresponding to 
the connected neighboring nodes are required to evaluate \eqref{init_KF_V} at node $n$ in the $k$-th iteration of 
our algorithm. 
To further elaborate, the proposed KF-DFPC algorithm operates in a distributed manner \cite{Xin_2022}, 
where the nodes run the Kalman filters on their own observations in Step $3(a)$ and $3(b)$ in Algorithm \ref{algo_2}, 
in parallel, and then locally broadcast the estimates and their error covariances to their immediate neighboring nodes 
to update the frequencies and phase across the array in Step $4$ and $5$. These locally shared values are 
then used to define the priors for the next iteration of KF using \eqref{KF_DFPC_upd} and \eqref{init_KF_V}.}

{The computational complexity of KF-DFPC in each iteration is dominated by equations 
\eqref{AC_eqns}, \eqref{time_upd_m}, and \eqref{time_upd_V}. Eqn. \eqref{AC_eqns} is part of the DFPC algorithm 
which has the computational complexity of $\mathcal{O}(\text{card}\{\chi_n\})$ per node, in which 
$\chi_n$ is the set of neighbors of node $n$ including itself, and the operation $\text{card}\{.\}$ 
computes the cardinality of this set. Eqns. \eqref{time_upd_m} and \eqref{time_upd_V} are 
part of KF which require inverting and then multiplying 
the $2\times 2$ matrices and thus has the computational complexity of $\mathcal{O}(8)$. 
Now since the KFs at all the nodes can be run in parallel, the computational complexity of KF-DFPC for 
each node per iteration is $\mathcal{O}(\text{card}\{\chi_n\}+8)$ which is the same as the KCIF and DKF algorithms 
proposed in \cite{Saber_2009, Sayed_2010}. Note that for the sparsely 
connected arrays with $\text{card}\{\chi_n\}\ll 8$, the computational complexity of KF-DFPC is 
$\mathcal{O}(8)$. In contrast, 
for the larger arrays with high connectivity and $\text{card}\{\chi_n\}\gg 16$, 
it becomes $\mathcal{O}(\text{card}\{\chi_n\})$. 
Thus, for the large densely connected arrays, 
the performance improvement by using KF with DFPC comes at no additional increase in the 
computational complexity.}
\subsection{Simulation Results}
{We evaluate the performance of the proposed KF-DFPC algorithm through the simulation results, and 
compare it to the DFPC algorithm, the 
diffusion LMS (DLMS) algorithm \cite{LMS_2010}, the Kalman consensus information filtering (KCIF) 
algorithm~\cite{Saber_2009}, and the diffusion Kalman filtering (DKF) algorithm~\cite{Xin_2022, Sayed_2010}.} 
We consider the same simulation set up as used in Section \ref{DFPC_sim} for the DFPC algorithm.  

{In Figs. \ref{fig:freq_vs_iter_KFDFPC} and \ref{fig:phase_vs_iter_KFDFPC}, we show the frequency and phase errors 
of all the $N=100$ nodes in the network vs. the number of iteration of KF-DFPC from a single trial when the 
connectivity between the nodes is 
$c=0.2$, $\text{SNR}=0$ dB, and the update interval is set as $T=0.1$ ms. It is observed that as the number 
of KF-DFPC iterations increase, both frequency and phase errors converge to the average of their initial values as expected. 
Furthermore, by comparing these figures with Figs. \ref{fig:freqs} and \ref{fig:phases}, we observe that KF-DFPC significantly 
reduces the residual phase error upon its convergence.}

Fig. \ref{fig:sigma_vs_N_KFDFPC} shows the standard deviation of the total phase errors 
of {the KF-DFPC, DLMS, KCIF, and DKF algorithms} upon convergence 
for different number of nodes $N$ in the network when two different connectivity values $c=0.2$ and $0.5$ 
are assumed. 
We set $\text{SNR}=0$ dB and the update interval as $T=0.1$ ms. 
The standard deviation of the total phase errors $(\sigma_\phi)$
are averaged over $10^3$ independent trials and we show the average value and the standard 
deviation of the samples using the error bar plot. 
We observe that for each value of $c$ as the number of nodes $N$ in the network increases the 
total phase error and its variation decreases 
{for all the algorithms. Furthermore, the model-free adaptive filtering algorithm, i.e., DLMS, results in 
the worst phase error among all the algorithms for both $c$ values 
because it does not take into account the state transitioning models of the nodes for computing the estimates. 
The KCIF algorithm fuses the measurements from the neighboring nodes to compute the MMSE estimates in each 
iteration, and thus it performs better than our proposed KF-DFPC and the DKF algorithms for the 
smaller $N$ values for both $c=0.2$ and $0.5$; however, as $N$ increases its $\sigma_\phi$ reaches a plateau 
due its suboptimal nature as described in \cite{Saber_2009}. At the larger $N$ values and higher $c$, 
the KF-DFPC and DKF algorithms perform better than KCIF due to the increase in the $D$ value which aids 
in computing more accurate frequency and phase average estimates per node, and due to the 
decrease in the second eigenvalue modulus $\lambda_2$ which reduces the residual phase error as shown 
in Section \ref{steady_state_section}.}
For larger values of $N$ the total 
phase error reaches an error floor resulting from the frequency and phase offset errors added per iteration. 
Furthermore, comparing Figs. \ref{fig:sigma_vs_N_DFPC} and \ref{fig:sigma_vs_N_KFDFPC} for $c=0.2$ and $\text{SNR}=0$ dB, 
it is observed that the KF-based algorithms significantly outperforms the DFPC algorithm for all $N$ values. 
Thus performing an MMSE estimation of 
the frequencies and phases at the nodes via Kalman filtering and then computing 
the averages certainly ensures a higher level of synchronization among the nodes. 

In Fig. \ref{fig:iter_vs_c_KFDFPC} we analyze the convergence speed of {the above algorithms} 
when the networks with different 
connectivity $c$ and different number of nodes are considered. In this figure, the convergence is 
declared when the standard deviation of the total phase error falls below {$\eta=0.1^\deg$ threshold 
which guarantees high coherence at the destination \cite{OCDA_2017}.} 
The observed convergence iterations are averaged over $10^3$ trials and we plot the average value 
and the standard deviation of the samples as before. 
{As expected, it is observed that 
for all $N$ number of nodes 
in the network, all algorithms converge faster with the increase in $c$, and for each $c$ having 
larger number of nodes $N$ results in a faster convergence. 
The DKF algorithm shows the worst convergence speed for all the $N$ and $c$ values as compared to other algorithms because 
it only fuses the MMSE estimates from the neighboring nodes 
in each iteration but does not fuse the error covariance matrices. On the other hand, 
our proposed KF-DFPC fuses the error covariance matrices as well, as described in \eqref{init_KF_V},
and thus converges significantly faster than DKF. For instance, for $c=0.05$ and $N=100$, DKF takes $30$ iterations on average 
whereas KF-DFPC takes just $2$ iterations. Note that the synchronization algorithm with faster convergence speed implies 
that a fewer number of messages needs to be exchanged between 
the nodes to reach synchronization, which reduces the energy consumption of the nodes and increases the 
lifetime of the system. 
This figure shows that our KF-DFPC algorithm also converges faster than the DLMS and KCIF algorithms for sparsely connected 
arrays. For e.g., for $c=0.1$ and $N=65$ nodes, KF-DFPC takes $8$ iterations, whereas KCIF takes $12$ iterations and DLMS takes $22$ iterations. 
By comparing Figs. \ref{fig:iter_vs_c_DFPC} and \ref{fig:iter_vs_c_KFDFPC}, it is observed that 
the KF-DFPC algorithm converges faster for the moderately connected 
networks with $c\in[0.05,0.8]$ for all the $N$ values than the DFPC algorithm. For instance, for $N=100$ nodes with 
network connectivity $c=0.05$, the KF-DFPC converges in $2$ iterations whereas DFPC required $16$ iterations, 
and when $N=65$ then the KF-DFPC converged in $4$ iterations whereas DFPC needed $158$ iterations. 
For the higher $c$ values, the required number of iterations for these algorithms are 
the same, however, a highly connected network can be challenging to implement in practice if the array is large. 
Thus, given the convergence characteristics and the smaller total phase error, 
the proposed KF-DFPC algorithm is preferable over the other algorithms for the synchronization in 
distributed arrays.}
\begin{figure}[tp]
        \centering
\includegraphics[width=0.50\textwidth]{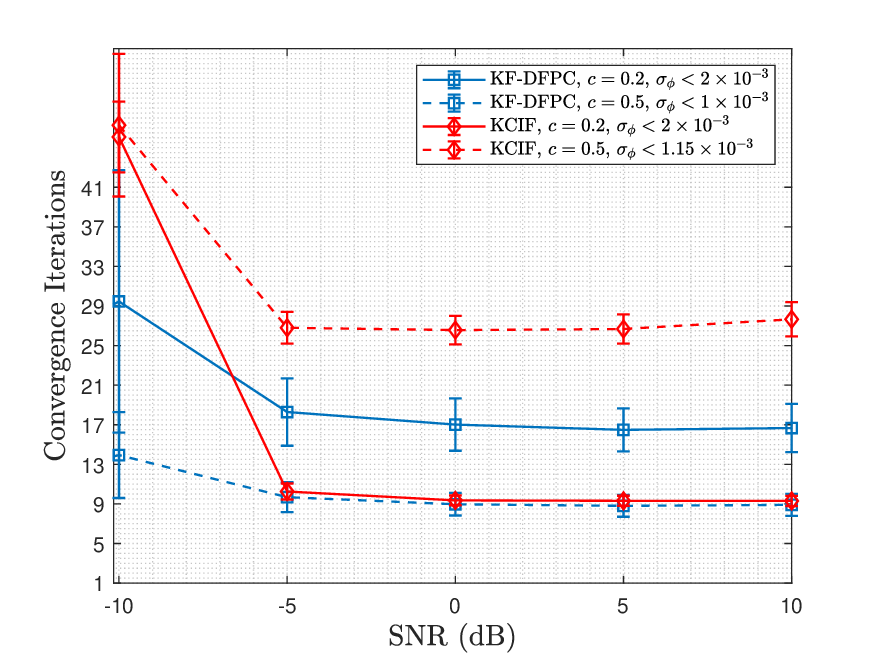}
	\caption{Convergence iterations of KF-DFPC and KCIF vs. $\text{SNR}$ with different connectivity levels $c$ 
	in the network and different thresholds on the final standard deviation of the total phase errors $\sigma_\phi$ 
	for $N=100$ and $T=0.1\text{ ms}$.}
	\label{fig:sigma_vs_SNR_KFDFPC}
\end{figure}

{Fig. \ref{fig:sigma_vs_SNR_KFDFPC} shows the convergence iterations of the KF-DFPC and KCIF algorithm as a function of 
$\text{SNR}$ when the network connectivity is $c=0.2$ or $0.5$. 
The number of nodes in the network is $N=100$ and we set the update interval as $T=0.1$ ms. 
The threshold on the 
standard deviation of the total phase error ($\sigma_\phi$) for each $c$ is selected  
based on the KF-DFPC and KCIF performances shown in Fig. \ref{fig:sigma_vs_N_KFDFPC}.
It is observed that for each $c$ value, both algorithms require more iterations to converge 
at the lower $\text{SNR}$ values, and as expected the required convergence iterations decreases 
with the increase $\text{in } c$. 
The increase in convergence iterations at the lower $\text{SNR}$s 
is because the Kalman filter used in these algorithms computes the MMSE estimates 
in each iteration by using the observation up to the present time instant (iteration), and thus with the decrease in $\text{SNR}$ 
more observations (iterations) need to be collected by KF to reduce the estimation error and 
reach the required total phase error. This implies that 
the KF-DFPC algorithm can achieve the total phase error irrespective of the $\text{SNR}$ in the system. 
This figure also shows that 
our KF-DFPC algorithm takes fewer iterations than KCIF at the lower $\text{SNRs}$ for the both $c$ values, 
and while for the sparsely connected array, KCIF shows faster convergence speed at the higher $\text{SNRs}$ due to the 
fusion of the neighboring nodes measurements at each node; nonetheless, as the connectivity between the nodes increases, 
the exchange of more measurements between the nodes in KCIF slow downs its 
convergence speed due to the independently varying states of the nodes and 
its use of a suboptimal Kalman gain~\cite{Saber_2009}. On the other hand, our KF-DFPC algorithm 
shows significant improvement than KCIF at the higher $c$ values as well.}

\begin{figure*}[tp]
		
    \begin{minipage}{0.48\textwidth}
        \centering
\includegraphics[width=0.99\textwidth]{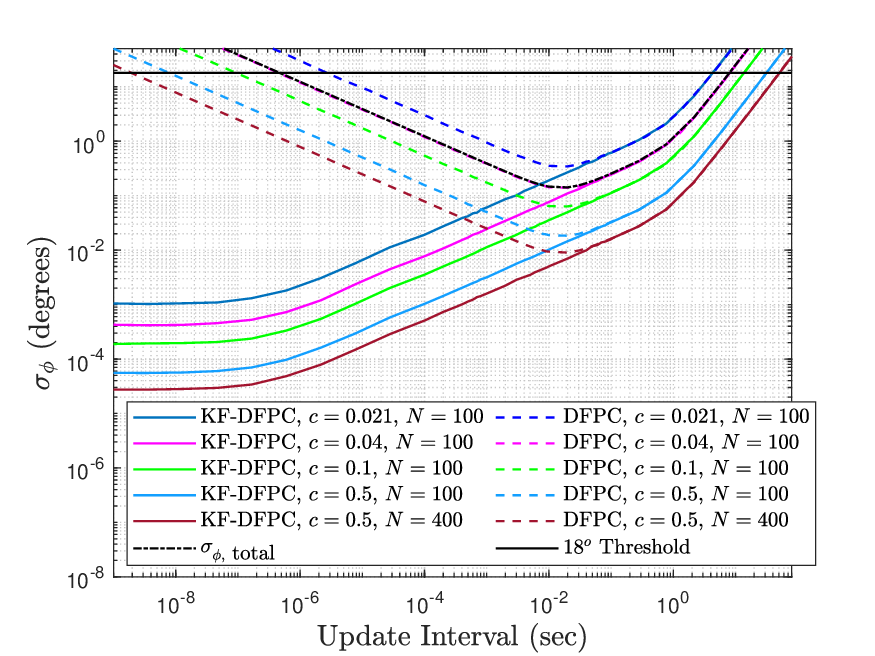}
	\caption{Standard deviation of the total phase error of KF-DFPC and DFPC algorithms vs. 
	the update interval $T$ for different connectivity $c$ and number of nodes $N$ when 
	$\text{SNR}=0$ dB.}
	\label{fig:sigma_vs_T}
		\end{minipage}\hspace{.025\linewidth}
    \begin{minipage}{0.48\textwidth}
        \centering
     	\includegraphics[width=0.99\textwidth]{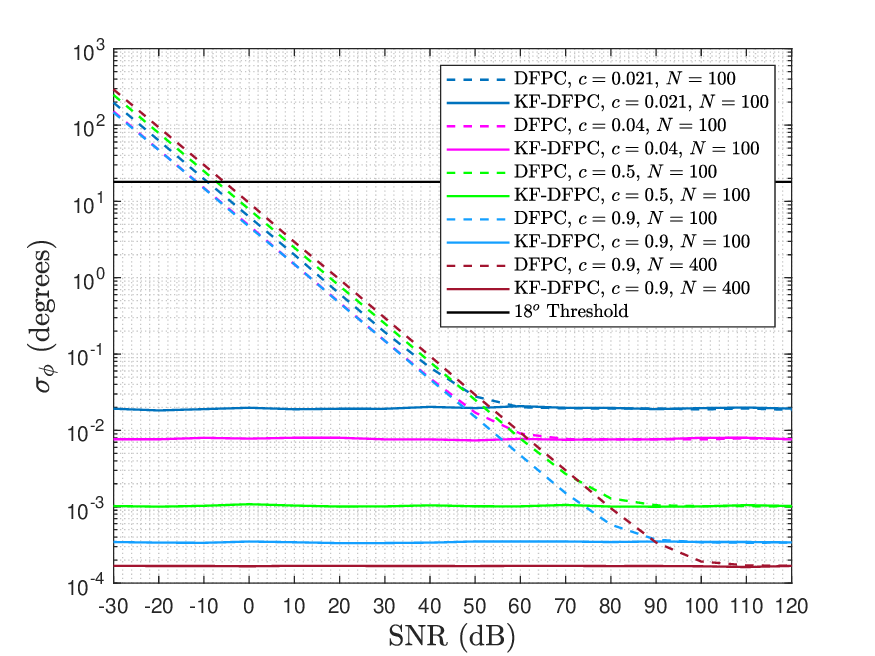}
	\caption{Standard deviation of the total phase error of KF-DFPC and DFPC algorithms vs. 
	signal to noise ratio $\text{SNR}$ for different connectivity $c$ and number of nodes $N$ when TDMA is used in the network 
	with $T=0.1$ ms.}
		\label{fig:sigma_vs_SNR_TDMA}
					\vspace{.2cm}
    \end{minipage}
\end{figure*}

\section{Residual Phase Error Evaluation}\label{residual_phase_section}
We compare the synchronization performance of the KF-DFPC and DFPC algorithms by analyzing the total phase error 
achieved upon convergence when varying the update interval $T$ and the signal to noise ratio $\text{SNR}$ of 
the signals. We assume that the nodes use 
a local broadcast to share their frequencies and phases with their neighboring nodes. 
Each node estimates the frequencies and phases of the signals 
by sampling them with sampling frequency $f_s=10$ MHz over the time duration $T$. 
The frequency drifts and phase jitters 
in the oscillators are modeled as described earlier in Section \ref{modeling_ftheta}. The standard deviation 
of the total phase errors are averaged over $200$ trials in the following figures.

In Fig. \ref{fig:sigma_vs_T}, we compare the standard deviation of the total phase error of the KF-DFPC and 
DFPC algorithm as a function of 
the update interval $T$ for different connectivity $c$ and different number of nodes $N$ in the network when 
$\text{SNR}=0$ dB is assumed. 
It is observed that for $T\in[0.1 \text{ms},3 \text{sec}]$, both DFPC and KF-DFPC algorithms 
yield total phase error below the $18^\deg$ threshold for the 
sparsely connected arrays with $c=0.021$ and the densely connected ones with $c\geq 0.5$. 
In particular, for each connectivity $c$, the DFPC algorithm gives a minimum total phase error at $T=20$ ms, but 
its total phase error increases for the update intervals above and below this value of $T$. This is because 
for the update intervals above $T=20$ ms, the errors due to the oscillators frequency drift dominate, and 
for the update intervals below this time duration, the errors due to the frequency and phase estimation are higher. 
Thus as expected the KF-DFPC algorithm outperforms the DFPC algorithm by large margins when $T\leq 20$ ms for each 
connectivity $c$ and every number of nodes $N$ in the network. 
For the purpose of analysis, in this figure 
we also plot the standard deviation of the phase error 
$\sigma_{\phi,\text{total}}$ as derived in \eqref{sigma_total}. 
As explained at the end of Section \ref{steady_state_section}, for the sparsely connected 
arrays, for instance, the arrays with $c=0.021$ as considered in this figure, 
since the second eigenvalue modulus $\lambda_2$ is close to $1$ and $\sum^I_{m=1}\lambda^{2m}_2\gg 1$, the 
total phase error upon the convergence of KF-DFPC for $T> 7$ ms and DFPC for all the $T$ values is 
higher than $\sigma_{\phi,\text{total}}$ bound. For connectivity $c=0.04$, we have 
$\sum^I_{m=1}\lambda^{2m}_2\approx 1$ and thus the total phase error from the DFPC follows the 
theoretical bound $\sigma_{\phi,\text{total}}$ for all the $T$ values, whereas this holds for the KF-DFPC 
algorithm only for $T>20$ ms as expected. As the connectivity $c$ increases further, the second eigenvalue 
modulus $\lambda_2$ approaches $0$ and thus the total phase error from both KF-DFPC and DFPC algorithms 
decreases below the $\sigma_{\phi,\text{total}}$ bound as seen from this figure.

Finally, in Fig. \ref{fig:sigma_vs_SNR_TDMA} we show the standard deviation of the total phase error 
$\sigma_\phi$ of the KF-DFPC and DFPC algorithms when varying the $\text{SNR}$ of the signals for the 
arrays with different connectivity $c$ and different number of nodes $N$. 
It is assumed that the nodes in the array use TDMA for communication, and thus 
the frequency and phase estimation errors are modeled using \eqref{freq_est_TDMA} and 
\eqref{phase_est_TDMA}. The update interval is set as $T=0.1$ ms. 
It is observed that for each connectivity $c$, the total phase error of the DFPC algorithm is higher 
at lower $\text{SNR}$ values but it decreases with the increase in the $\text{SNR}$ and reaches the phase 
error of the KF-DFPC algorithm at the higher $\text{SNR}$ values. 
This performance degradation of DFPC at 
lower $\text{SNR}$s is because the residual error due to frequency and phase estimation is higher 
which as result increases its total phase error upon convergence. This estimation error is also higher for 
connectivity $c=0.9$ when $N=400$ vs. when $N=100$ due to the decrease in the time duration of sampling.
However, higher $\text{SNR}$ is in practice difficult to achieve in the systems 
because it requires increasing the transmitted signal power which is limited due to the hardware 
constraints. Since the KF-DFPC algorithm 
computes the MMSE estimates of the frequencies and phases in each iteration, it maintains the 
same total phase error irrespective of the $\text{SNR}$ of the signals. Note that this consistent phase error 
of KF-DFPC is realized at the cost of a few additional convergence iterations at the lower $\text{SNR}$ values 
as shown earlier in Fig. \ref{fig:sigma_vs_SNR_KFDFPC}. 

\section{Conclusions}\label{conclusion_section}

A decentralized approach to jointly synchronizing the frequencies and phases of separate nodes in a distributed antenna array was presented. Based on local broadcast of the electrical states of each node, consensus averaging supports convergence to within a residual phase error commensurate with high coherent beamforming gain.
We independently modeled the frequency drifts and phase jitters of the oscillators and the 
frequency and phase estimation errors at the nodes using practical statistics. 
A decentralized algorithm (DFPC) computing a weighted average of the frequencies and phases of its neighboring nodes was analyzed, where the phases and frequencies of the shared signals were modeled with estimation errors. 
Simulation 
results showed that the DFPC algorithm synchronizes the nodes up to a non-negligible residual phase error 
that results from the frequency and phase offset errors at the nodes, but that under certain conditions this residual phase error is below that needed to support high coherent beamforming gain.
Furthermore, DFPC takes a large 
number of iterations to converge for moderately connected arrays with fewer number of nodes,
which increases the energy consumption of the array and introduces delay in achieving synchronization. 
Although its synchronization performance improves at the larger $\text{SNR}$s, in practice the change in $\text{SNR}$ 
is limited by the hardware constraints and is also dependent on multipath fading in communication channels 
between the nodes. Thus to reduce the residual phase error irrespective of the $\text{SNR}$ 
and improve the synchronization between the nodes, 
a Kalman filtering based decentralized algorithm (KF-DFPC) was also proposed. The KF-DFPC algorithm 
reduces the residual phase error by computing 
the MMSE estimates of the frequencies and phases at the nodes before computing the weighted avenges. 
{The synchronization performance of KF-DFPC was compared to the DFPC, DLMS, KCIF, and DKF algorithms.
Simulation results demonstrate that the KF-DFPC algorithm 
significantly reduces the residual phase error upon convergence at short update 
intervals as compared to the DFPC and DLMS algorithms. 
In addition, under certain conditions, KF-DFPC converges in fewer iterations than all these algorithms}, 
and its synchronization performance is independent of the $\text{SNR}$ of the received signals.

\bibliographystyle{IEEEtran}
\bibliography{References}
\end{document}